\documentclass[titlepage,11pt]{article}

\usepackage[myheadings]{fullpage}
\usepackage[pdftex]{graphicx}

\usepackage{pmetrika}
\usepackage{pmbib}
\usepackage{submitMio}

\usepackage{amsmath,amssymb,latexsym}

\usepackage{url}

\usepackage{setspace}
\usepackage{graphicx} 
\usepackage{array} 
\usepackage{bm} 
\usepackage{booktabs} 
\usepackage{ctable} 
\usepackage{subfig} 
\usepackage{caption} 
\usepackage{footnote}
\usepackage{threeparttable}
\usepackage{color}
\usepackage{rotating}
\usepackage{lscape}
\usepackage{tensor}
\usepackage[linewidth=1.5pt]{mdframed}
\usepackage{xcolor,colortbl}
\usepackage{pifont} 
\usepackage[round]{natbib}
\makeatletter 
\renewcommand\@biblabel[1]{#1.} 
\makeatother %

\definecolor{Gray}{gray}{0.85}

\graphicspath{{Figures/}} 

\DeclareMathOperator{\Exp}{Exp}
\DeclareMathOperator{\Gam}{Ga}

\DeclareMathOperator{\Dir}{Dir}

\DeclareMathOperator{\PP}{\mathbf{P}}

\DeclareMathOperator{\uomega}{\underline{\omega}}

\DeclareMathOperator{\upi}{\underline\pi}

\usepackage{enumerate}
\usepackage{xr}
\begin{document}

%
%
%
%
%
%
%
%
%

\setcounter{page}{1}
\vspace*{2\baselineskip}

\RepeatTitle{Bayesian Plackett-Luce mixture models for partially ranked data}\vskip3pt

\linespacing{1.2}

\author{Cristina Mollica and Luca Tardella}

\affil{Dipartimento di Scienze statistiche, Sapienza Universit\`a di Roma, Rome, Italy} 
\vspace{.2cm}
\contact{The final publication is available at Springer via \url{http://dx.doi.org/10.1007/s11336-016-9530-0}}

\abstracthead

\begin{abstract}

The elicitation of an ordinal judgment on multiple alternatives is often required in many psychological and behavioral experiments to investigate preference/choice orientation of a specific population.
The Plackett-Luce model is one of the most popular and frequently applied parametric distributions to analyze rankings of a finite set of items. 
The present work introduces a Bayesian finite mixture of Plackett-Luce models
to account for unobserved sample heterogeneity of partially ranked data. We describe an efficient way to incorporate the latent group structure in the data augmentation approach and the derivation of existing maximum likelihood procedures as special instances of the proposed Bayesian method.
Inference can be conducted with the combination of the Expectation-Maximization algorithm for maximum \textit{a posteriori} estimation and the Gibbs sampling iterative procedure.
We additionally investigate several Bayesian criteria for selecting the optimal mixture configuration 
and describe diagnostic tools
for assessing the fitness of
ranking distributions conditionally and unconditionally on the number of ranked items.
The utility of the novel Bayesian parametric Plackett-Luce mixture for characterizing sample heterogeneity is illustrated with several
applications to simulated and real preference ranked data. We compare our method with the frequentist approach and a Bayesian nonparametric mixture model both assuming the Plackett-Luce model as a mixture component.
Our analysis on real datasets reveals the importance of an accurate diagnostic check for an appropriate in-depth understanding of the heterogenous nature of the partial ranking data.

\end{abstract}
\vspace{\fill}\pagebreak


\section{Introduction}
\label{s:intro}
Choice behavior is a theme of great interest in several research areas, such as social and psychological sciences,
but its investigation usually involves variables which cannot be directly observed and measured in an objective and precise manner.
For this reason, the evidence in choice experiments is often collected in ordinal form, that is, in terms of \textit{ranking data}.
More specifically, ranked data arise in those studies where a sample of $N$ people is presented a finite set of $K$ alternatives, called \textit{items}, and is asked to rank them according to a certain criterion, such as personal preferences or attitudes. Thus, a generic ranking is the result of a comparative judgment on the competing alternatives expressed in the form of order relation. Interest in ranked data analysis is motivated, for example, by marketing and political surveys,
but also by psychological and behavioral studies consisting, for instance, in the ordering of
words/topics according to the perceived association with a reference subject.

Ranked data analysis has been addressed from numerous perspectives, as revealed by a wide and consolidated literature reviewed in \cite{Marden} and, more recently, in \cite{Alvo}.
Of course, a significant role is played by
the parametric modeling of ranking data, which sometimes 
is inspired by possible patterns underlying the (random) mechanism of 
formation of individual preferences. 
Nowadays, there is a large number 
of parametric ranking distributions but, 
despite the large availability of options, 
often none of them 
are able to embody 
the appropriate flexibility 
to represent the heterogeneous nature of real data. 
Consequently, it is natural to extend them 
to the
mixture context.
Our work focuses on 
the finite mixture approach 
with the Plackett-Luce model (PL) as a parametric component
within a Bayesian inferential framework, aimed at analyzing heterogeneous partial rankings.
It parallels the frequentist approach in 
\cite{Gormley:Murphy-Royal}.
Recent works considering Bayesian mixture modeling based on the PL are
\cite{Gormley:Murphy-BayesianAnalysis}
 and \cite{Caron:Teh2014}.
\cite{Gormley:Murphy-BayesianAnalysis} 
deal with a grade of membership model
where, at each stage of the sequential ranking process,
each sample unit has a specific partial membership of each component.
This model is inherently different from the usual finite mixture model with discrete distributions on the latent variable developed in \cite{Gormley:Murphy-Royal} and is better suited for soft clustering purposes.
A Bayesian nonparametric PL based on a Gamma process to account for an infinite number of items, shortened as BNPPL, is developed in \cite{Caron:Teh:Murphy2012}. The BNPPL has been subsequently extended to the mixture context, hereinafter abbreviated as BNPPLM, by \cite{Caron:Teh2014} for analyzing clustered partial ranking data.  
This work relies on modeling the exchangeable sequence of random partial orderings 
with an infinite mixture derived by means of a 
stick-breaking construction of the weights, corresponding to a Dirichlet process mixture. 
Hence, although one can consider the BNPPLM developed by \cite{Caron:Teh2014} as a natural generalization of our finite mixture framework, we point out two important differences: 
(i) in our parametric setting, each single component is a standard PL for finite orderings (possibly truncated), whereas
the BNPPL component models the orderings of a possibly arbitrary number of items; 
(ii) in our framework, the cardinality of the mixture models is explicitly defined as finite, whereas it is infinite in \cite{Caron:Teh2014}. Hence, the ability of the BNPPLM to identify a suitable finite number of clusters underlying the observed data is related to the random partition associated to the sequential draw of partial rankings. In fact, for each sample unit the partial ranking is generated from the corresponding random vector of support parameters, which in turn follows a    
Dirichlet allocation model \citep{Mccullagh2008}.
Multiple sample units can then share the same parameter vector and, hence, belong to the same group of the partition.
One can rely on the posterior simulation 
of the parameters and use, as suggested by  \cite{Caron:Teh2014}, the \textit{ad hoc} method originally proposed by \cite{Dahl} to estimate a suitable finite number of underlying groups. 

In order to address the typical issues faced with a parametric finite mixture analysis, 
we devote special attention to 
alternative criteria for the  determination of the 
appropriate number of components. 
Additionally, we investigate suitable diagnostic tools to detect possible deficiencies of the PL
parametric class 
in capturing the underlying 
dependence structure and 
highlight some critical issues in combining partial orderings characterized by a different number of ranked items.
Indeed, we will show how this step is relevant for an appropriate recognition of the 
parsimonious group structure.

The outline of the article is the following. In Section~\ref{s:pl}
we review 
the PL for partial orderings and its Bayesian estimation based on data augmentation.
The novel Bayesian PL mixture and the related inferential procedures are presented in Section~\ref{s:bmpl}, 
together with  
alternative Bayesian model selection criteria 
and model assessment diagnostics.
Illustrative applications of the proposed methods to both simulated and
real ranking data are presented in Section~\ref{s:app}. 
In Section~\ref{s:concl} the paper ends with
concluding remarks and hints to future developments.


\section{The Plackett-Luce model}
\label{s:pl}

\subsection{Model specification}
\label{ss:modspec}

A ranking can be elicited through a series of sequential comparisons in which a single item is preferred to all the remaining alternatives and, after being selected, is removed from the next comparisons.
This is the basic construction underlying the PL, a well-established parametric distribution among the so-called \textit{stagewise ranking models}. 
It was originally  
introduced by~\cite{Luce} and~\cite{Plackett}.
More specifically, by denoting with $K$ the total number of items to be ranked, 
the PL
is parametrized by the 
\textit{support parameters} $\underline{p}=(p_1,\dots,p_K)$
representing positive constants associated to each item:
the higher the value of the support parameter $p_i$,
the greater the probability for the $i$-th item to be preferred at each selection stage.
Let
$\underline{\pi}^{-1}=\{\pi_s^{-1}\}_{s=1}^N$
be a random sample
consisting of $N$ partial top
orderings
of the form $\pi^{-1}_s=(\pi^{-1}_s(1),\dots,\pi^{-1}_s(n_s))$.
With a slight abuse of notation,
$n_{s}$ is the length of the $s$-th partial ordering, that is, the number of items ranked by unit $s$ in the top $n_{s}$ positions.
The remaining $K-n_s$ items are assumed to be ranked lower.
In our notation, a full ordering corresponds to the case $n_s=K-1$ 
since, once $K-1$ items have been ranked, the last position is automatically determined.
Under the PL the contribution to the likelihood from the $s$-th partial ordering is given by
%
\begin{equation}
\label{e:pl}
\PP_{\text{PL}}(\pi^{-1}_s|\underline{p})=\prod_{t=1}^{n_s}\dfrac{p_{\pi_s^{-1}(t)}}{\sum_{i=1}^{K}p_i-\sum_{\nu=1}^{t-1}p_{\pi_s^{-1}(\nu)}}.
\end{equation}
We notice that for strictly partial orderings ($n_s<K-1$) the distribution in
~\eqref{e:pl} corresponds to the 
marginal PL distribution for full orderings obtained by integrating out the items ranked in the last $K-n_s$ positions. 
An important summarizing feature of $\PP_{\text{PL}}(\cdot|\underline{p})$ is the \textit{modal ordering} $\sigma^{-1}_{\underline{p}}$, corresponding to the ordering of the support parameters $\underline{p}$ from the largest to the smallest. 

\subsection{Model estimation}
\label{ss:modest}
The main inferential issue related to formulation \eqref{e:pl}
concerns 
the presence of the annoying normalization term
$\left(\sum_{i=1}^{K}p_i-\sum_{\nu=1}^{t-1}p_{\pi_s^{-1}(\nu)}\right)$ that does not permit the
direct maximization of the likelihood. In the maximum likelihood estimation (MLE) framework,
\cite{Hunter} overcomes this difficulty by applying the
Minorization-Maximization algorithm, an iterative optimization method relying
on the replacement of the original PL log-likelihood with a minorizing
surrogate objective function. In the Bayesian perspective, instead, a
related efficient solution is derived by \cite{Caron:Doucet},
whose work can be considered the starting point of our parametric proposal presented in the next section.
In particular, \cite{Caron:Doucet} propose to introduce a data augmentation step with latent quantitative variables $\underline{y}=(y_{st})$ for $s=1,\dots,N$ and $t=1,\dots,n_s$, 
whose conditional joint distribution is given by
\begin{equation}
\label{e:fullcond}
f(\underline{y}|\upi^{-1},\underline{p})=\prod_{s=1}^N\prod_{t=1}^{n_s}f_{\Exp}\biggl(y_{st}\biggr\vert\sum_{i=1}^{K}p_i-\sum_{\nu=1}^{t-1}p_{\pi_s^{-1}(\nu)}\biggr),
\end{equation}
where $f_{\Exp}(\cdot|\lambda)$ denotes the Negative Exponential density
with rate parameter
$\lambda$.
The parametric assumption \eqref{e:fullcond}
entails remarkable simplifications
for the implementation of both the posterior optimization and the Gibbs Sampling (GS) algorithm.
The success of the Bayesian device introduced by \cite{Caron:Doucet} is due to the combination of \eqref{e:fullcond} with a conjugate prior specification.
This latter aspect moves from the \textit{Thurstonian interpretation} of \eqref{e:pl}, that is, Thurstone's ranking model 
reduces to the PL when the Gumbel distribution is employed as distribution of the latent scores; see \cite{Yellott}.
\cite{Caron:Doucet} exploited the conjugacy of the Gamma density with the Gumbel distribution and derived a simple and effective GS scheme for the approximation of the posterior distribution.

 \section{Bayesian mixture of Plackett-Luce models}
\label{s:bmpl}
A wide variety of research contexts require a model-based analysis
accounting for the presence of differential patterns in a collection of partially ranked data.
To our knowledge, Bayesian inference of a finite PL mixture has not been previously developed in the literature concerning parametric methods to analyze such data. 
Bayesian PL estimation appeared so far in the literature is either limited to the homogeneous case, as in \cite{Guiver:Snelson} and \cite{Caron:Doucet}, or accounts simultaneously for an infinite mixture configuration and an infinite number of items through a nonparametric approach; see \cite{Caron:Teh:Murphy2012,Caron:Teh2014}.
In the next subsections, we detail the novel Bayesian PL mixture model for partial top rankings.

\subsection{Model and prior specification}
\label{ss:ms}
Let
$\underline{\pi}^{-1}$ be a random sample of partial top orderings with varying lengths drawn from a $G$-component PL mixture,
in symbols
\begin{equation*}
\label{e:mpl}
\pi_1^{-1},\dots,\pi_N^{-1}|\underline{p},\uomega \overset{iid}{\sim}\sum_{g=1}^G\omega_g\PP_{\text{PL}}(\pi^{-1}_s|\underline{p}_g),
\end{equation*}
where $\underline{p}_g$ is the support parameter vector specific of the $g$-th mixture component and $\omega_g$ is the corresponding weight.
In order to suitably generalize the data augmentation approach in~\cite{Caron:Doucet} within the finite mixture framework, we need to introduce an additional latent feature of each generic sample unit $s$, represented by the unobserved group labels
\begin{equation*}
\underline{z}_s=(z_{s1},\dots,z_{sG})|\uomega\overset{iid}{\sim}\text{Multinom}(1,\uomega=(\omega_1,\dots,\omega_G)),
\end{equation*}
whose univariate marginal distribution corresponds to a Bernoulli random variable (r.v.) such that
\begin{equation*}
z_{sg}=\begin{cases}
      1\qquad \text{if unit $s$ belongs to the $g$-th mixture component}, \\
      0\qquad \text{otherwise}.
\end{cases}
\end{equation*}
%
We propose to include the unobserved group labels $\underline{z}$ in the data augmentation strategy as follows:
\begin{equation}
\label{e:y}
f(\underline{y}|\upi^{-1},\underline{z},\underline{p},\uomega)=\prod_{s=1}^N\prod_{t=1}^{n_s}f_{\Exp}\left(y_{st}\biggr\vert\prod_{g=1}^G\left(\sum_{i=1}^{K}p_{gi}-\sum_{\nu=1}^{t-1}p_{g\pi_s^{-1}(\nu)}\right)^{z_{sg}}\right).
\end{equation}
%
This implies that the latent group labels determine the cluster-specific support parameters acting on the underlying quantitative variables $\underline{y}$.
Once the model governing observed and latent variables is specified, a fully Bayesian approach requires the elicitation of the joint prior distribution for the unknown parameters.
We choose prior distributions
with independent $\underline{p}$ and $\uomega$, so that $f_0(\underline{p},\uomega)=f_0(\underline{p})f_0(\uomega)$, and a convenient conjugate structure, similarly to the homogeneous population case.
For the support parameters, in fact, we extend the initial distribution in~\cite{Caron:Doucet} by defining independent $p_{gi}\sim\Gam(c_{gi},d_g)$,
where the Gamma r.v.'s are indexed by the shape and the rate parameter.
Finally, for the mixture weights, taking values in the $(G-1)$-dimensional simplex, we make the standard prior assumption $\uomega\sim\Dir(\alpha_1,\dots,\alpha_G)$.

\subsection{MAP estimation}
\label{ss:mapest}

In the presence of the latent variables $\underline{y}$ and $\underline{z}$,
we can construct an EM algorithm in order to optimize the posterior distribution and learn
the posterior mode
(MAP estimate).
The complete-data likelihood can be factorized as $L_c(\underline{p},\uomega,\underline{y},\underline{z})
=f(\underline{y}|\upi^{-1},\underline{z},\underline{p},\uomega)\PP(\upi^{-1},\underline{z}|\underline{p},\uomega)$,
that is, the product of the full-conditional \eqref{e:y} times the
standard complete-data likelihood of a mixture model specification
without data augmentation with $\underline{y}$.
With simple algebra, 
both factors of the complete-data likelihood
can be rearranged in order to explicit a multinomial form in $\underline{z}$
as follows:
\begin{equation*}
f(\underline{y}|\upi^{-1},\underline{z},\underline{p},\uomega)
=\prod_{s=1}^N\prod_{g=1}^G
\left(\prod_{t=1}^{n_s}\left(\sum_{i=1}^{K}p_{gi}-\sum_{\nu=1}^{t-1}p_{g\pi_s^{-1}(\nu)}\right)e^{-\sum_{t=1}^{n_s}y_{st}\left(\sum_{i=1}^{K}p_{gi}-\sum_{\nu=1}^{t-1}p_{g\pi_s^{-1}(\nu)}\right)}\right)^{z_{sg}}
\end{equation*}
and
\begin{equation*}
\PP(\upi^{-1},\underline{z}|\underline{p},\uomega)
=\prod_{s=1}^N\prod_{g=1}^G\left(\omega_g\prod_{t=1}^{n_s}\dfrac{p_{g\pi_s^{-1}(t)}}{\sum_{i=1}^{K}p_{gi}-\sum_{\nu=1}^{t-1}p_{g\pi_s^{-1}(\nu)}}\right)^{z_{sg}}.
\end{equation*}
Hence,
\begin{equation*}
\label{}
L_c(\underline{p},\uomega,\underline{y},\underline{z})
=\prod_{s=1}^N\prod_{g=1}^G\left(\omega_g\prod_{i=1}^Kp_{gi}^{u_{si}}
e^{-p_{gi}\sum_{t=1}^{n_s}\delta_{sti}y_{st}}
\right)^{z_{sg}},
\end{equation*}
where
\begin{equation*}
u_{si}=\begin{cases}
      1\quad\text{ if }i\in\{\pi_s^{-1}(1),\dots,\pi_s^{-1}(n_s)\},\\
      0\quad\text{ otherwise},
\end{cases}
\end{equation*}
and
\begin{equation*}
\delta_{sti}=\begin{cases}
      1\quad\text{ if }i\notin\{\pi_s^{-1}(1),\dots,\pi_s^{-1}(t-1)\},\\
      0\quad\text{ otherwise},
\end{cases}
\end{equation*}
with $\delta_{s1i}=1$ for all $s=1,\dots,N$ and $i=1,\dots,K$.
%
We denote the complete-data log-likelihood with $l_c(\underline{p},\uomega,\underline{y},\underline{z})=\log L_c(\underline{p},\uomega,\underline{y},\underline{z})$. 
%
%
The implementation of the EM algorithm in the Bayesian framework
iterates (i) the M-step, maximizing with respect to $(\underline{p},\uomega)$ the following objective function:
%
\begin{equation*}
\label{e:VEM}
Q((\underline{p},\uomega),(\underline{p}^*,\uomega^*))=\mathbb{E}_{\underline{y},\underline{z}|\upi^{-1},\underline{p}^*,\uomega^*}[l_c(\underline{p},\uomega,\underline{y},\underline{z})]+\log f_0(\underline{p},\uomega);
\end{equation*}
%
(ii) the E-step, which relies on the conditional joint distribution of all the latent variables given by
\begin{equation*}
\label{e:startVEM}
\PP(\underline{y},\underline{z}|\upi^{-1},\underline{p},\uomega)=f(\underline{y}|\upi^{-1},\underline{z},\underline{p},\uomega)
\PP(\underline{z}|\upi^{-1},\underline{p},\uomega).
\end{equation*}
%
The E-step returns
\begin{equation*}
\begin{split}
Q((\underline{p},\uomega),(\underline{p}^*,\uomega^*))
&=\sum_{s=1}^N\sum_{g=1}^G \hat z_{sg}\biggl(\log\omega_g+\sum_{i=1}^K\biggl(u_{si}\log p_{gi}-p_{gi}\sum_{t=1}^{n_s}\dfrac{\delta_{sti}}{\sum_{i=1}^{K}\delta_{sti}p^*_{gi}}\biggr)\biggr)\\
&\qquad+\sum_{g=1}^G(\alpha_g-1)\log\omega_g+\sum_{g=1}^G\sum_{i=1}^K\bigl((c_{gi}-1)\log p_{gi}-d_gp_{gi}\bigr),
\end{split}
\end{equation*}
where the posterior membership probabilities $\hat z_{sg}$ are obtained as 
\begin{equation*}
\label{}
\hat z_{sg}=\dfrac{\omega_g^*\PP_{\text{PL}}(\pi^{-1}_s|\underline{p}_g^*)}{\sum_{g'=1}^G\omega_{g'}^*\PP_{\text{PL}}(\pi^{-1}_s|\underline{p}_{g'}^*)}.
\end{equation*}
%
Differentiating the objective function $Q$ with respect to\ each $p_{gi}$
and equating to zero yields the updated support parameters of the M-step
%
\begin{equation*}
\label{}
p_{gi}=\dfrac{c_{gi}-1+\hat\gamma_{gi}}{d_g+\sum_{s=1}^N \hat z_{sg}\sum_{t=1}^{n_s}\dfrac{\delta_{sti}}{\sum_{i=1}^{K}\delta_{sti}p^*_{gi}}}
\end{equation*}
for $g=1,\dots,G$ and $i=1,\dots,K$, where $\hat\gamma_{gi}=\sum_{s=1}^N \hat z_{sg}u_{si}.$
Optimizing $Q$ with respect to\ $\uomega$, subject to the canonical constraint
$\sum_{g=1}^G\omega_g=1$,
yields the updated mixture weights
\begin{equation*}
\label{e:weightMAP}
\omega_g=\dfrac{\alpha_g-1+\sum_{s=1}^N \hat z_{sg}}{\sum_{g'=1}^G\alpha_{g'}-G+N}\qquad g=1,\dots,G.
\end{equation*}
%
Notice that when $G=1$ the 
MAP
procedure collapses into the single updating formula obtained by \cite{Caron:Doucet}.
Moreover, similarly to their method, also in our mixture approach we can recover the MLE as special case of the noninformative Bayesian analysis with flat priors,
obtained
by setting $c_{gi}=1$, $d_g=0$ and $\alpha_g=1$. Such a configuration of the hyperparameters, in fact, reduces
the proposed MAP estimation
to the algorithm described 
by~\cite{Gormley:Murphy-Royal}
in the 
frequentist framework.

\subsection{Gibbs Sampling}
\label{ss:gsest}

In order to draw a sample from the joint posterior distribution and learn about the uncertainty associated to the final estimates, we detail the implementation of a GS procedure.
The conjugate prior configuration described in Section~\ref{ss:ms}, combined with the complete-data likelihood $L_c(\underline{p},\uomega,\underline{y},\underline{z})$,
leads
to a sampling scheme with simple parametric distributions to be drawn from.
In particular, the full-conditionals of the latent component labels are easily derived by noting that $\PP(\underline{z}|\upi^{-1},\underline{y},\underline{p},\uomega)\propto
L_c(\underline{p},\uomega,\underline{y},\underline{z}),$
implying the following multinomial structure:
\begin{equation*}
\PP(\underline{z}_s|\pi_s^{-1},\underline{y}_s,\underline{p},\uomega)\propto
\prod_{g=1}^G\left(\omega_g\prod_{i=1}^Kp_{gi}^{u_{si}}
e^{-p_{gi}\sum_{t=1}^{n_s}\delta_{sti}y_{st}}
\right)^{z_{sg}}.
\end{equation*}
%
The full-conditionals of the support parameters
are still members of the Gamma family
with hyperparameters suitably updated as follows:
\begin{equation*}
\PP(p_{gi}|\upi^{-1},\underline{y},\underline{z},p_{[-gi]},\uomega)\propto
f_0(p_{gi})L_c(\underline{p},\uomega,\underline{y},\underline{z})
\propto p_{gi}^{c_{gi}+\gamma_{gi}-1}e^{-p_{gi}(d_g+\sum_{s=1}^Nz_{sg}\sum_{t=1}^{n_s}\delta_{sti}y_{st})},
\end{equation*}
where
%
$\gamma_{gi}=\sum_{s=1}^Nz_{sg}u_{si}$
%
%
is the number of units belonging to cluster $g$ who have ranked item $i$ and
$p_{[-gi]}$ denotes the matrix $\underline{p}$ of the support parameters without the $(g, i)$-th entry.
Also the full-conditional of the mixture weights
has the same form of
the corresponding prior class,
obtained as
\begin{equation*}
\PP(\uomega|\upi^{-1},\underline{y},\underline{z},\underline{p})
\propto f_0(\uomega)L_c(\underline{p},\uomega,\underline{y},\underline{z})
\propto f_0(\uomega)\PP(\underline{z}|\uomega)
= \prod_{g=1}^G\omega_g^{\alpha_g+\sum_{s=1}^Nz_{sg}-1}.
\end{equation*}
%
Finally,
the full-conditional of $\underline{y}$ is given by construction
in the assumption \eqref{e:y}. 

Note that the EM and the GS can be conveniently combined by employing the MAP solution as good initialization of the chain in the MCMC simulation.
However, 
when one adopts an MCMC procedure to derive Bayesian approximate inference of a mixture model,
the MCMC sample can be affected by the annoying identifiability issue, known as
\textit{label switching} phenomenon (LS). This may
 prevent from a straightforward posterior estimation \citep{Celeux:Hurn:Robert,Marin:Mengersen:Robert}.
In the Bayesian PL mixture applications presented in Section~\ref{s:app} we exploited alternative relabeling algorithms that perform an \textit{ex post} rearrangement of the raw MCMC drawings in order to 
obtain meaningful posterior estimates. These were implemented by means of the functions included in the recently released R package \texttt{label.switching} \citep{Papastamoulis}.

\subsection{Determining the number of components}
\label{s:baymodel}
In the estimation procedures previously described, the number $G$ of
groups is fixed \textit{a priori}.
Thus, after performing a separate
inference on PL mixtures with a different number of components,
a method for discriminating among the competing models is needed. 
In our applications, we explored three types of alternative Bayesian criteria to address this issue: (i) \textit{Deviance Information Criterion} (DIC) introduced by \cite{Spiegelhalter}, (ii) \textit{Bayesian Information Criterion-Monte Carlo} (BICM) proposed by \cite{Raftery2007} and (iii) \textit{Bayesian Predictive Information Criterion} (BPIC) described in \cite{Ando}.
For an updated detailed review of Bayesian tools for model
comparison, see \cite{Gelman:Hwang:Vehtari}.
We start from the general formula $\text{DIC}=\bar D+p_{D}$, where $\bar D=\mathbb{E}[D(\theta)|\upi^{-1}]$ is the posterior expected deviance with $D(\theta)=-2\log L(\theta)$ and $p_{D}$ represents the \textit{effective number of parameters}.
We
consider two alternative DIC formulations corresponding to two alternative ways of conceiving $p_D$, i.e.,
$\text{DIC}_1=\bar D+(\bar D-D(\hat\theta_\text{MAP}))$, based on the MAP estimate $\hat\theta_\text{MAP}$, and 
$\text{DIC}_2=\bar D+\mathbb{VAR}[D(\theta)|\upi^{-1}]/2$, suggested by \cite{Gelman}. 
As shown in \cite{Raftery2007}, DIC$_2$ coincides with AICM, that is, the Bayesian counterpart of AIC. We also use two versions
of BICM, specifically
$\text{BICM}_1=\bar D +
\frac{\mathbb{VAR}[D(\theta)|\upi^{-1}]}{2}(\log N -1)$, which is
based on the approximation of the MAP estimate from the MCMC sample
\citep{Raftery2007}, and $\text{BICM}_2=D(\hat\theta_\text{MAP}) +
\frac{\mathbb{VAR}[D(\theta)|\upi^{-1}]}{2}\log N$. 
Finally, since one aspect often debated on DIC is its tendency to overfit, due to the double usage of the observed data, we additionally employ two BPIC formulations
obtained from DIC$_1$ and DIC$_2$ 
by doubling their 
penalty term $p_{D}$. 

After fitting mixture models with alternative number of components, one can select a suitable number $\hat G$ of components 
using a specific criterion and identifying the optimal mixture 
which minimizes that criterion. 
Since alternative criteria can lead to different choices, 
we will compare and discuss the possibly different selections of optimal models and provide some recommendation based on a simulation study.

\subsection{Model assessment}
\label{s:baymodcheck}

Once the optimal PL mixture model has been selected, a comprehensive inferential analysis should also contemplate
the adequacy of the estimated model in describing the observed data
\citep{Gelman:Meng:Stern}.
In this regard, we have focused on two important features of the 
ranking data $\underline{\pi}^{-1}$: 
\begin{enumerate}[(i)] %
\item the most liked item frequency vector
$\underline{r}(\upi^{-1})$, whose generic entry $r_i(\upi^{-1}) =\sum_{s=1}^N I_{[\pi^{-1}_{s}(1)=i]}$ 
counts
how many times item $i$ 
is ranked first;
\item
the paired comparison frequency matrix 
$\tau(\upi^{-1})$,
whose generic entry
%
\begin{equation*}
\begin{split}
\tau_{ii'}(\upi^{-1})=\sum_{s=1}^N(1-(1-u_{si})(1-u_{si'}))I_{[\pi_s(i)<\pi_s(i')]}
=\sum_{s=1}^N(u_{si}+u_{si'}-u_{si}u_{si'})I_{[\pi_s(i)<\pi_s(i')]}
\end{split}
\end{equation*}
%
counts the number of times that item $i$ is preferred to item $i'$.
\end{enumerate}
Within the Bayesian paradigm, 
it is possible to
generalize the classical goodness-of-fit statistic into a parameter-dependent quantity, 
referred to 
as \textit{discrepancy variable} \citep{Gelman:Meng:Stern},
and perform a posterior predictive check of model goodness-of-fit.
Let us denote with $\upi^{-1}_\text{obs}$ the observed collection of partial
orderings and with $\upi_{\text{rep}}^{-1}$ a replicate random  
draw from the posterior predictive distribution
under the specified model $H$.
The \textit{posterior predictive $p$ value} 
based on a generic discrepancy variable
$X^2(\upi^{-1};\theta)$
is defined as 
\begin{equation}
\label{e:postp}
p_B=\PP\left(X^2(\upi_{\text{rep}}^{-1};\theta)\geq X^2(\upi^{-1}_\text{obs};\theta)\biggr\vert\upi^{-1}_\text{obs},H\right).
\end{equation}
%
Under correct model specification, $p_B$ is expected to be 
close to 0.5, whereas small values 
are deemed as an indication of model
inadequacy. 
Here we considered 0.05 as critical threshold.

Indeed, as a first type of $X^2$ discrepancy measure we have considered 
$$
X^2_{(1)}(\upi^{-1};\theta) = \sum_{i=1}^K 
\frac{\left(
r_i(\upi^{-1}) - r_i^*(\theta)
\right)^2}
{r_i^*(\theta)},
$$
where the symbol $^*$ indicates the theoretical frequency expected under PL mixture model with parameter $\theta=(\underline{p},\uomega)$.
By following \cite{Yao}, as a second discrepancy measure 
we have considered
$$X_{(2)}^2(\upi^{-1};\theta)=\sum_{i<i'}\frac{(\tau_{ii'}(\upi^{-1})-\tau^*_{ii'}(\theta))^2}{\tau^*_{ii'}(\theta)}.$$
Details on 
the computation of the posterior
predictive $p$ values, denoted with 
$p_{B(d)}$ $d=1,2$ and
corresponding to each 
discrepancy measure $X_{(d)}^2$, 
are reported in the
Supplementary Material (SM). 



Moreover, whenever partial ranking data show considerable proportions of strictly
partial rankings with a different number of ranked items,
one can
further investigate model adequacy conditionally on the observed length of the partial orderings.
This 
can help to verify possible violations of the
underlying assumption that the subsets of rankers are identically
distributed, such that their preference system is driven by the same mixture
distribution on the support parameters.
This check could reveal that one should better account for sample heterogeneity. 
To this aim, 
we have defined two other discrepancy measures, 
$\tilde{X}_{(1)}^2$ and $\tilde{X}_{(2)}^2$, which parallel the
previous ones.
The corresponding 
posterior predictive $p$ values,
denoted with 
$\tilde{p}_{B(d)}$ $d=1,2$, 
allow to assess the homogeneity assumption of the
strata of rankers characterized by different lengths 
of the expressed partial orderings. Details are reported in the SM.

\section{Illustrative applications}
\label{s:app}

We will apply our Bayesian model to simulated as well as two real datasets. We will verify its comparative performance with respect to some natural alternative methods, which can be expected to perform similarly, and highlight some possible advantages. We first provide some implementation details.  
Although the Bayesian approaches described in Section~\ref{ss:mapest} and \ref{ss:gsest}
permit to convey specific (subjective) prior knowledge on the parameters, in the following analyses we will rely upon weakly/noninformative prior densities
with hyperparameters equal to $c_{gi}=1$, $d_g=.001$ and $\alpha_g=1$, in order to
allow also for a direct comparison with the frequentist PL mixture developed by~\cite{Gormley:Murphy-Royal}.
For our parametric method, 
we first recorded the MAP estimate derived through the EM algorithm 
and subsequently employed it to initialize the GS. We run the MCMC algorithm for a total of 22000 iterations and discarded the first 2000 drawings as burn-in period. 
Moreover, the application of the alternative relabeling algorithms 
on the MCMC posterior samples revealed a good performance in removing the LS and returned very similar results in terms of adjusted estimates. Posterior means were used as final parameter estimates and those reported for the considered experiments were derived, specifically, with the application of the pivotal reordering algorithm \citep{Marin:Mengersen:Robert}.

In assessing the performance of our method, we will focus 
also on the comparison with 
the BNPPLM,
since it 
represents 
the most recent natural competitor to handle heterogeneity of partial ranking data and
can be expected to perform similarly.
In the BNPPLM analysis, we employed the default setup for the hyperparameters described in \cite{Caron:Teh2014} and run their GS algorithm for 100000 iterations.

\subsection{Simulation study}
\label{s:appsimnew}

\begin{table}[t]
\caption{{\em Simulation study} Percentages of top$-m$ partial orderings for the three censoring settings considered in the simulation study.}
\small 
\label{t:censoringrates}
\centering
\begin{tabular}{cccccc}
 & \multicolumn{5}{c}{$m$} \\ 
\cline{2-6}  
Censoring &  &  &  &  &  \\
setting  & 1 & 2 & 3 & 4 & 5  \\
\hline
A & 0 & 2 & 4 & 10 & 84   \\
B & 5 & 15 & 15 & 20 & 45  \\
C & 5 & 20 & 20 & 25 & 30  \\
 \hline
\end{tabular}
\end{table}
%

We considered a simulation plan with four different PL mixture population scenarios, where the true number $G^*$ of 
components ranges from 1 
to 4.
Specifically, in Scenario $c\in\{1,\dots,4\}$ 
one has $G^*=c$. 
Under each scenario, we simulated 100 samples composed of $N=1000$ complete orderings of $K=6$ items. 
The values of the support parameters for each dataset were randomly generated as $p_{gi}\overset{iid}{\sim}\text{Beta}(0.3,0.3)$, where the U-shaped Beta density aims at guaranteeing a sufficient separation among the mixture components.
Additionally, we assumed equal weights by setting $\omega_g=1/G^*$ for all $g=1,\dots, G^*$ and $G^*=1,\dots,4$. 
In order to perform the analysis on partial observations, a censoring was randomly induced on the complete orderings.
In each scenario, we separately considered three censoring settings (A, B and C) for the random truncation of the complete data: the percentages of the number $m$ 
of top ranked items are detailed in Table~\ref{t:censoringrates}. In censoring setting A,
the percentages of partial orderings with the same number of ranked items 
were set equal to those
observed in the CARCONF data considered in subsection \ref{s:apprealcar}. This yields approximately 16\% of strictly partial orderings in each simulated sample. Censoring settings B and C are characterized by increasing proportions of truncation yielding, respectively, 
55\% and 70\% of strictly partial observations. In this way, we are able to thoroughly explore the effectiveness of our parametric framework and 
its sensitivity to 
differential presence of strictly partial rankings in the sample.
Bayesian finite PL mixtures, with a number $G$ of components ranging from 1 to 7, and the BNPPLM were fitted to all the artificial datasets 
for each population scenario and censoring setting. 
The comparison between the two models was based on the
performance regarding the 
identification of the actual number $G^*$ of groups in the four scenarios.
In our Bayesian parametric PL mixture analysis, the optimal number $\hat G$ of groups
was identified by means of the alternative model selection criteria described in subsection \ref{s:baymodel}. Tables~\ref{t:Distr1},~\ref{t:Distr2} and~\ref{t:Distr3} show the distribution of $\hat G$ for the alternative criteria as well as
that corresponding to the BNPPLM analysis obtained with the optimization method of \cite{Dahl}, as suggested by \cite{Caron:Teh2014}. More briefly, Table~\ref{t:agreement} displays only the agreement rates (\%).
Regarding censoring setting A, in Scenario 1
BPIC$_1$, BPIC$_2$ and BICM$_1$ always recover the actual absence of 
heterogeneity (i.e. $G^*=1$). On the other hand, this happens also for the frequentist approach employing BIC as well as for the BNPPLM.
For the remaining population scenarios, BPIC$_1$ and DIC$_1$ perform from slightly to substantially better, especially in the case $G^*=4$ where DIC$_1$ emerges with an agreement rate of 81\%, followed by BPIC$_1$ with 77\%. 
The gap with both the frequentist and nonparametric results is considerable.
BIC exhibits an agreement rate equal to 66\%, whereas for BNPPLM the rate is remarkably smaller. In fact, only for 50\% of the datasets BNPPLM fitted  
$\hat G=G^*$ components.
With the application of 
censoring settings B and C on the same synthetic data,
we faced with the situation when most of the sequences to be analyzed are partial. With both censoring settings, for $G^*<4$ the results associated to the Bayesian rules and BIC were found to be substantially robust with respect to 
censoring setting A, and DIC$_1$ and BPIC$_1$ still differ in 
the best agreement rates. In the case $G^*=4$, instead, the negative effect of the higher truncation percentage becomes more evident. In fact, we noted an overall worsening of the performance of all the selection methods, especially for the BNPPLM. Nonetheless, similarly to 
censoring setting A, 
DIC$_1$ exhibits the highest agreement rates (76\% and 72\%), confirming its sizable advantage over BIC (57\% and 55\%) and the BNPPLM (50\% and 37\%).
Apparently, in almost all cases the BNPPLM approach yields the lowest agreement rates.
Moreover, in the presence of heterogeneity ($G^*>1$), BNPPLM is consistently associated with the greatest variability regarding the determination of the number $\hat G$ of groups;
see the corresponding distributions in Tables~\ref{t:Distr1},~\ref{t:Distr2} and~\ref{t:Distr3}.
If on one hand the relatively worse performance of BNPPLM is partly due to the fact that data are simulated from a different generative model, on the other it highlights 
a possibly substantial difference between the two approaches.
Overall, our simulation study suggests 
to privilege 
the use of BPIC$_1$ and DIC$_1$ for the subsequent analysis, 
although with a higher occurrence of partial observations DIC$_1$ seems to be slightly preferable.

We conclude the simulation study by providing some evidence on the fitting measures presented in subsection~\ref{s:baymodcheck}. For succinctness, only results for the computation of $p_{B(d)}$ $(d=1,2)$ 
under the most critical population scenario with $G^*=4$ components 
are shown. 
Boxplots of $p_{B(d)}$ values for all the parametric PL mixtures fitted to the 100 simulated datasets 
are reported in Figure~\ref{fig:Diagnostic} as a function of the number $G$ of fitted components. They point out the effectiveness of the proposed diagnostic tools to highlight possible deficiencies of misspecified PL mixture models. As expected, we observe an increasing trend of the posterior predictive $p$ values over $G<4$, whereas for $G\geq 4$ they stabilize around the reference value 0.5.

\begin{table}[t]
\caption{{\em Simulation study} Agreement rates (\%) between true number $G^*$ of PL components in the four population scenarios and the optimal number $\hat G$ of components identified, respectively, by the Bayesian PL mixture analysis via alternative model selection criteria and by the BNPPLM analysis via Dahl's procedure. Best agreement rates for each simulation scenario and censoring setting are highlighted in bold.}
\small 
\label{t:agreement}
\centering
\resizebox{0.7\textwidth}{!}{  
\begin{tabular}
 {lcccccccc}
\multicolumn{9}{c}{Censoring setting A} \\
\hline 
$G^*$ & DIC$_1$ & DIC$_2$ & BPIC$_1$ & BPIC$_2$ & BICM$_1$ & BICM$_2$ & BIC & BNPPLM \\
\hline
1 & 99 & 98 & \bf{100} & \bf{100} & \bf{100} & 99 & \bf{100} & \bf{100} \\
2 & 96 & 93 & \bf{98} & 95 & 93 & 89 & 97 & 91\\
3 & 91 & 89 & \bf{94} & 92 & 91 & 88 & 93 & 88\\ 
4 & \bf{81} & 67 & 77 & 70 & 65 & 60 & 66 & 50\\ 
\multicolumn{9}{c}{Censoring setting B} \\
\hline
$G^*$ & DIC$_1$ & DIC$_2$ & BPIC$_1$ & BPIC$_2$ & BICM$_1$ & BICM$_2$ & BIC & BNPPLM \\
\hline 
1 & 99 & \bf{100} & \bf{100} & \bf{100} & \bf{100} & 98 & \bf{100} & \bf{100} \\ 
2 & 97 & 93 & 97 & \bf{98} & 95 & 89 & 95 & 89 \\ 
3 & 92 & 88 & \bf{94} & 92 & 90 & 82 & 92 & 79 \\ 
4 & \bf{76} & 61 & 69 & 65 & 53 & 48 & 57 & 50 \\ 
\multicolumn{9}{c}{Censoring setting C} \\
\hline
$G^*$ & DIC$_1$ & DIC$_2$ & BPIC$_1$ & BPIC$_2$ & BICM$_1$ & BICM$_2$ & BIC & BNPPLM \\
\hline 
1 & 97 & 98 & 98 & \bf{100} & \bf{100} & \bf{100} & \bf{100} & \bf{100} \\ 
2 & \bf{97} & 91 & \bf{97} & 95 & 91 & 87 & 95 & 90\\ 
3 & \bf{95} & 89 & 94 & 92 & 88 & 82 & 92 & 63\\ 
4 & \bf{72} & 62 & 67 & 64 & 48 & 46 & 55 & 37 \\ 
\hline
\end{tabular}
}
\end{table}

\subsection{The CARCONF data}
\label{s:apprealcar}

Our second analysis concerns a marketing study described in \cite{Dabic:Hatz:2009}, aimed at investigating customer preferences toward different car features.
The car configurator (CARCONF) dataset consists of $N=435$ top orderings and is available in the R package \texttt{prefmod} \citep{Hatz:Ditt:2012}.
Customers were asked to construct their car by using an
online configurator system.
The respondents were presented a set of $K=6$ car modules to carry out their personal preferences, namely 1=price, 2=exterior design, 3=brand, 4=technical equipment, 5=producing country, and 6=interior design.
The survey did not require a complete ranking elicitation; therefore the sample is composed of partial top orderings. The distribution of the varying number of ranked items
%
\begin{figure}[t]
\centering
\includegraphics[scale=.5]{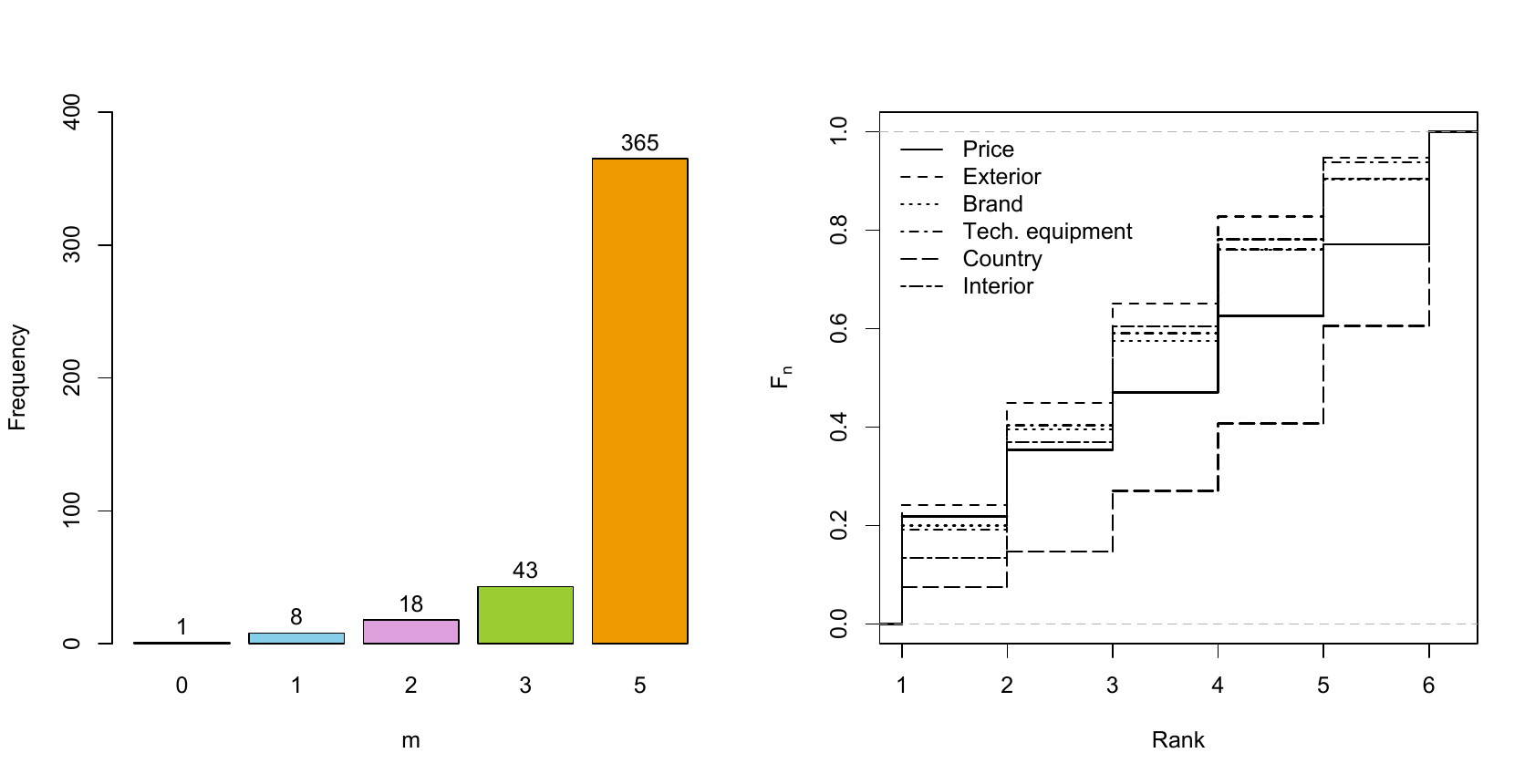}
\caption
{{\em CARCONF data} Distribution of the number $m$ of ranked items (\textit{left}) and empirical c.d.f.'s of the marginal rank distributions of the car features (\textit{right}).}
\label{f:freqmissCAR}
\end{figure}
%
is detailed in Figure \ref{f:freqmissCAR} (left).
Most of the customers (365 units, 84\% of the sample) submitted a complete ordering
of the car features.
Most of the remaining customers
submitted a strictly partial ordering by providing their top-4 favorite features.
The vector (42,17,0,29,62,27) lists the number of missing responses for each item.
Hence, all respondents assigned a rank to the brand, whereas the producing country is the one whose exact position is more frequently missing (62 occurrences corresponding to 89\% of the total number of incomplete responses).
The producing country is also associated with the lowest mean rank, as indicated by the fifth entry of the average rank vector $\bar\pi=(3.56,2.88,3.17,3.11,4.49,3.20)$.
The graphical inspection of the marginal rank distribution for each item, reported in the form of empirical c.d.f. in Figure \ref{f:freqmissCAR} (right), provides additional details on the overall preferences.
We note that the c.d.f. for the producing country is remarkably stochastically dominated by the other ones, matching the idea of a minor global interest in the car production country.
Another important aspect to be highlighted is the presence of intersections among the c.d.f.'s that can be interpreted as an empirical violation to the assumption of an underlying homogeneous PL, under which the rank distributions are instead expected to be \textit{marginally stochastically ordered} \citep{Marden}. The observed behavior of the rank distributions could be explained with the existence of differential preference patterns in the sample. These patterns could be better captured by assuming an underlying group structure with an unknown number of groups, rather than with a basic homogeneous PL.

\begin{figure}[t]
\centering
\subfloat{ 
\includegraphics[scale=.34]{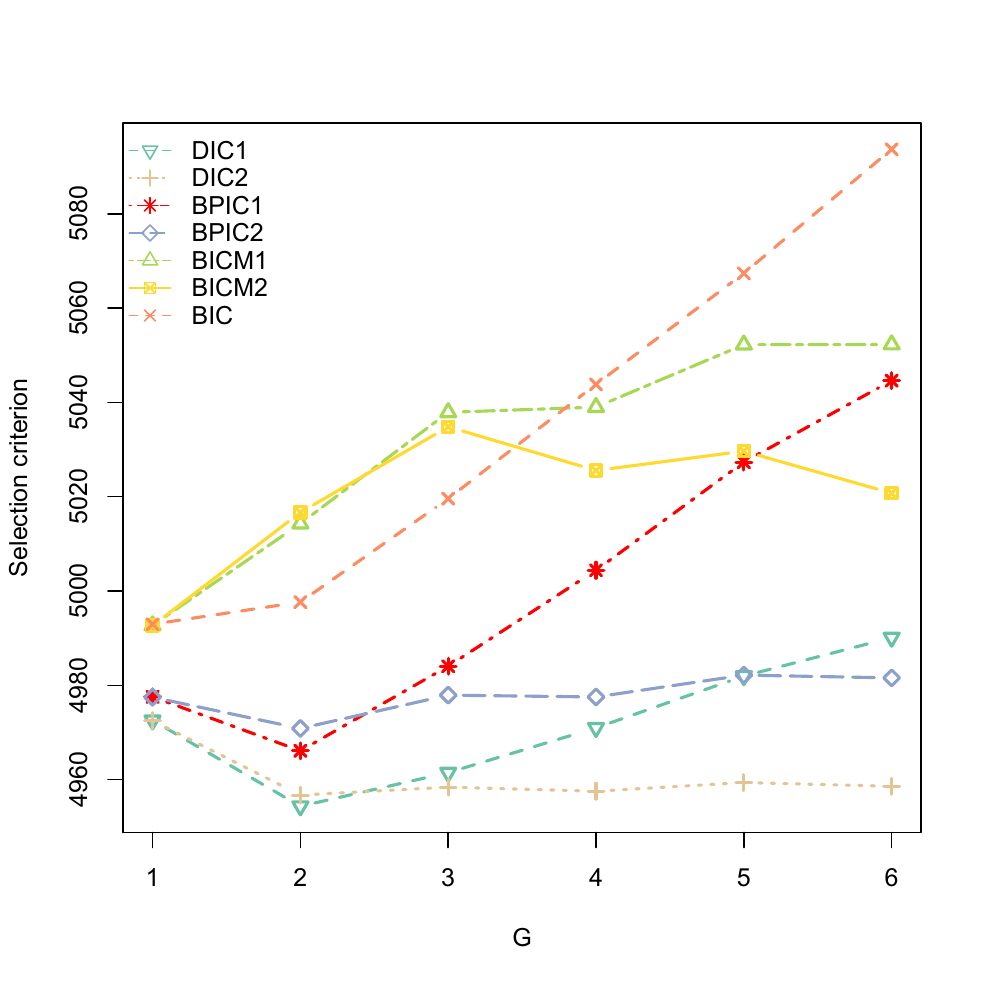}
\label{fig:CAR1}}
\subfloat{ 
\includegraphics[scale=.36]{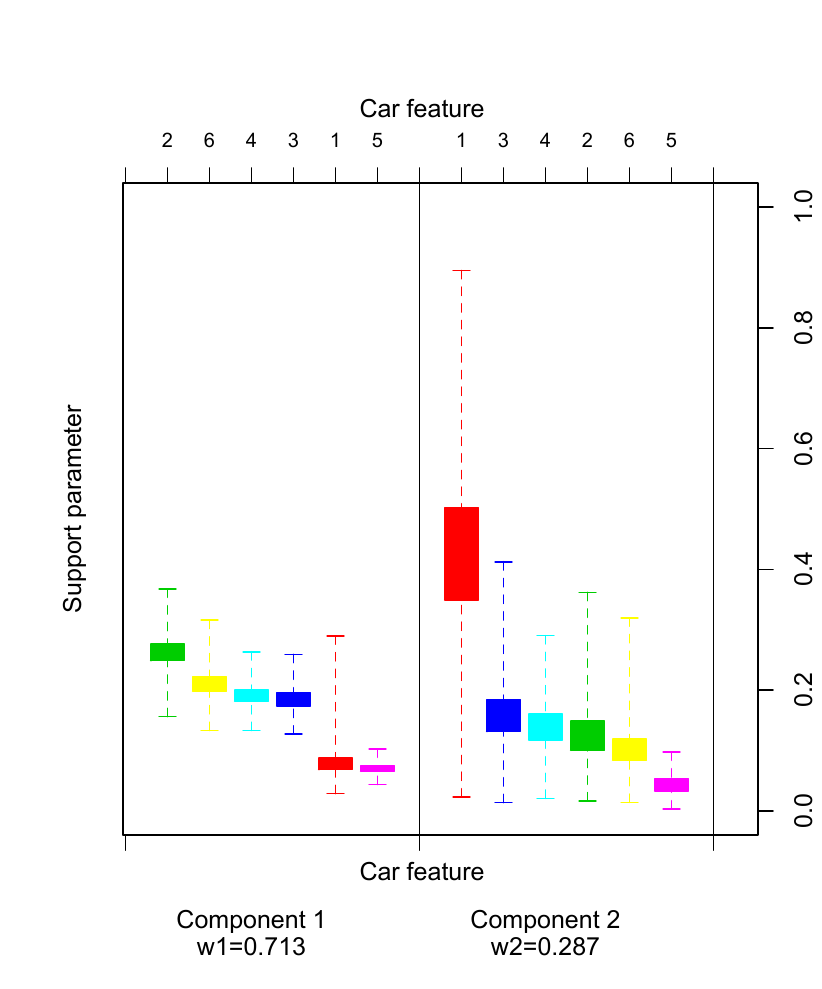}
\label{fig:CAR1}}
\subfloat{ 
\includegraphics[scale=.35]{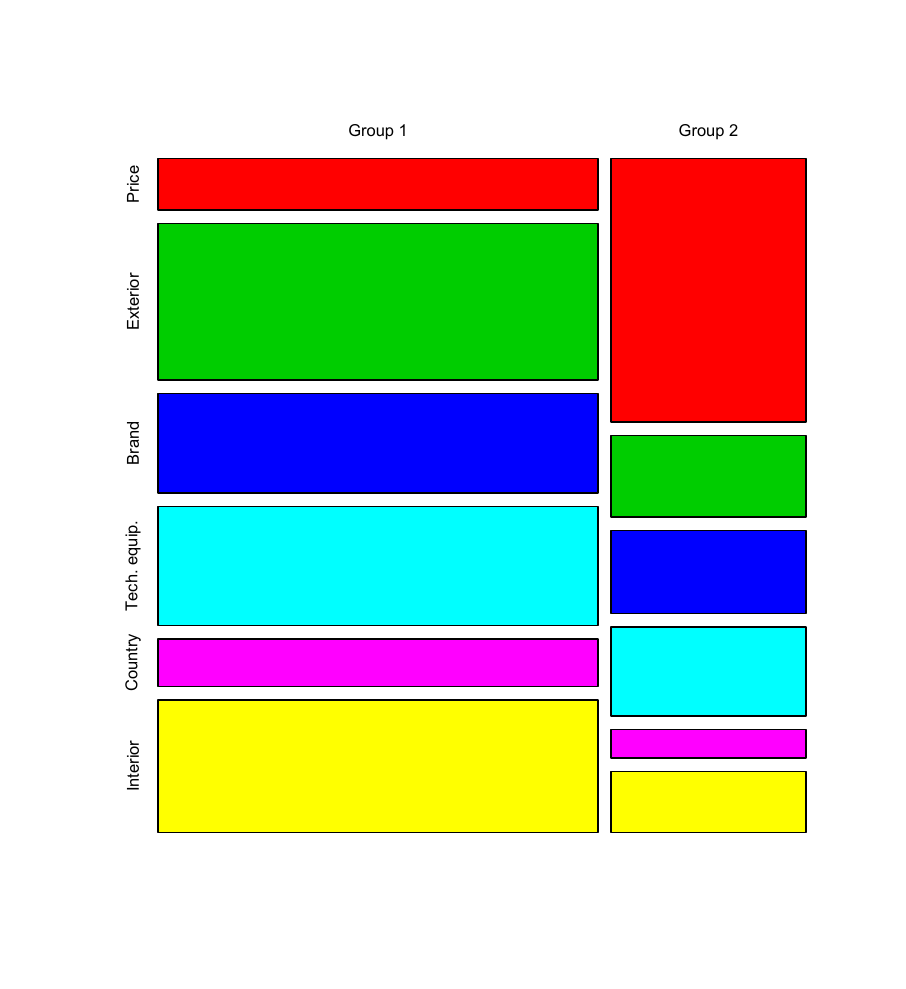}
\label{fig:MosaicCar}}
\caption{{\em CARCONF data} Model selection criteria (\textit{left}) for the Bayesian PL mixtures with a varying number $G$ of components. For each selection criterion, the optimal choice of the number of components corresponds to the minimum value. Boxplots (\textit{center}) and mosaic plot (\textit{right}) for the posterior samples of the support parameters of the optimal Bayesian 2-component PL mixture.}
\label{fig:Selec-MosaicCar}
\end{figure}

We estimated PL mixtures on the CARCONF data with a number of components varying from $G=1$ to $G=6$.
Bayesian selection criteria and BIC are shown in Figure~\ref{fig:Selec-MosaicCar} (left). Numerical details 
are in Table \ref{t:Baycrit3}. 
BIC, as well as 
BICM$_1$ and BICM$_2$, does not recognize the existence of an underlying group structure despite the large sample size, whereas all versions of DIC and BPIC agree in selecting the 2-component PL mixture.
The application of the BNPPLM to the CARCONF dataset
agrees with the MLE inference identifying 
a single PL component with 
vector of estimated support parameters equal to $\hat{\underline{p}}=(0.123,0.231,0.195,0.193,0.071,0.187)$.
In fact, the homogeneity conclusions of the \textit{ad hoc} criterion 
in~\cite{Dahl} 
matches with the degenerate 
distribution of the number of distinct support parameter vectors associated to each unit across the posterior simulations. This result conflicts somehow with our preliminary descriptive findings on the violation of the stochastic dominance among the marginal rank distributions.

For all the fitted Bayesian PL mixtures, posterior predictive $p$ values are reported in the last two columns of Table \ref{t:Baycrit3}. 
The posterior predictive $p$ value $p_{B(2)}=0.505$
highlights a good fit in terms of the ability of the model to reproduce the bivariate features related to the pairwise comparisons, whereas $p_{B(1)}=0.079$ 
reveals a possible deficiency of the model to recover the marginal probability of the most favorite item, although it is larger than the usual 0.05 critical threshold. On the other hand, we notice that $p_{B(1)}<10^{-4}$ is well below for $G=1$, supporting the need of a heterogeneous model.

Parameter estimates of the optimal 2-component PL mixture are displayed
in Figure \ref{fig:Selec-MosaicCar} (center and right) and detailed in Table \ref{t:carest}.
The selected mixture model suggests the presence of a major cluster ($\hat\omega_1=0.713$)
comprising customers
mainly interested in 
esthetic features, with greater support to the exterior ($\hat p_{12}=0.263$) rather than interior design ($\hat p_{16}=0.211$). The minor group ($\hat\omega_2=0.287$), instead, is characterized by a greater attention in the economic aspect represented by the price ($\hat p_{21}=0.436$).
Both groups share a minor interest in the production country ($\hat p_{15}=0.071$ and $\hat p_{25}=0.043$). These results seem to better accord with the typical preference patterns observed in the car market than the homogeneous scenario.

%
%

\subsection{The APA data}
\label{s:apprealapa}

Another interesting dataset involving partial rankings is the popular 1980 American Psychological Association (APA) election dataset. The entire APA dataset with $N=15449$ voters ranking a maximum of $K=5$ candidates is available in the R package \texttt{ConsRank}. 
%
\begin{figure}[t]
\centering
\includegraphics[scale=.5]{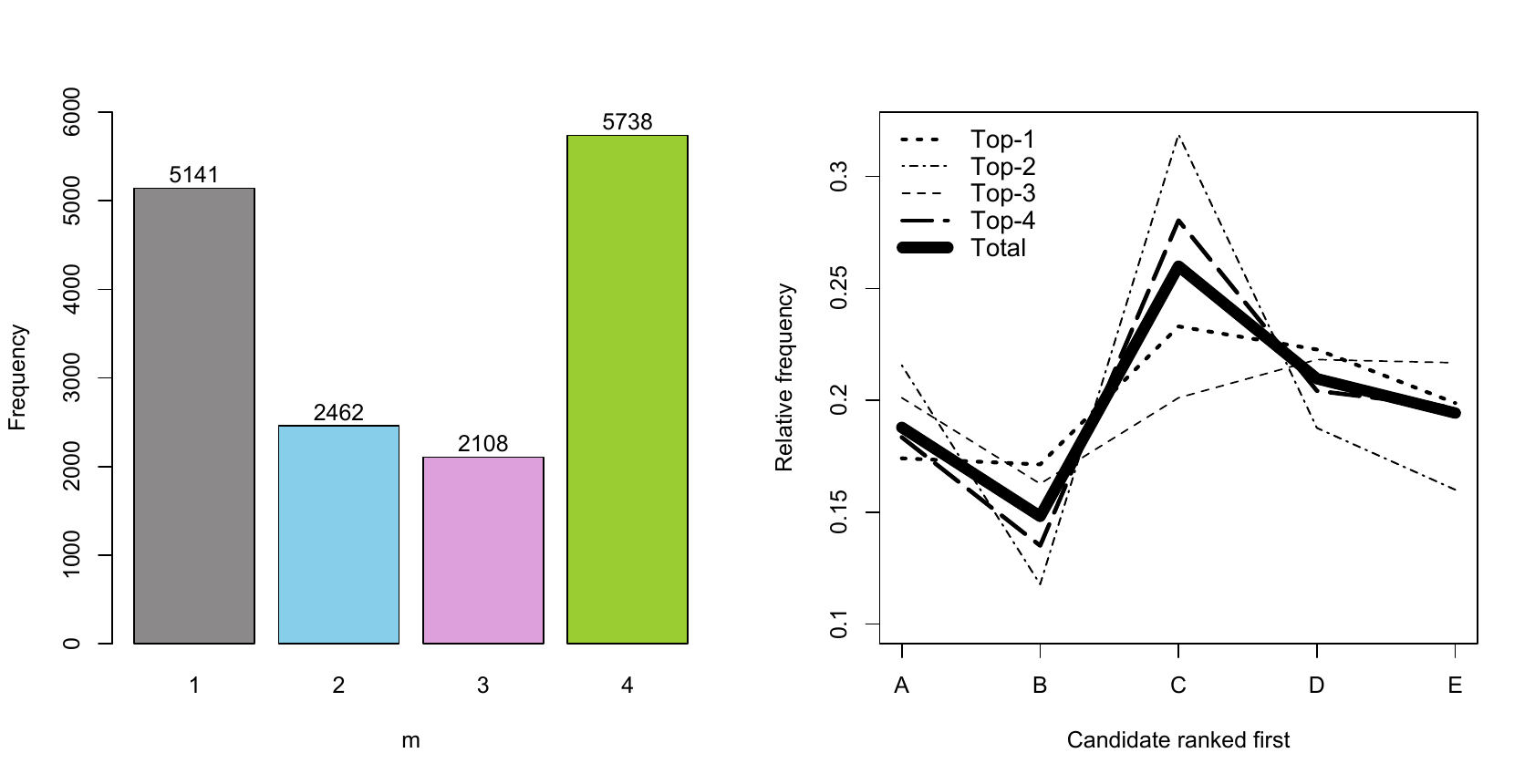}
\caption
{{\em APA election data} Distribution of the number $m$ of ranked items (\textit{left}) and distribution of the candidate occupying the first position by number of ranked items (\textit{right}).}
\label{f:freqmissAPA}
\end{figure}
%
The majority of the ballots (9711, 63\%) contain strictly partial orderings of the most favorite candidates and in most of them (5141, 33\%)
just a single favorite candidate is recorded. Some descriptive statistics are shown in Figure~\ref{f:freqmissAPA} and in the SM (Table~\ref{t:firstorder} and~\ref{t:firstchoice}). A detailed explanation of the data collection and the corresponding voting system yielding the elected candidate can be found in \cite{Diaconis-Rep}.

The popularity of these data is testified by the numerous attempts to provide an account of the complex heterogeneous structure of the ballots. From the pioneering descriptive analysis by \cite{Diaconis-Rep}, relying on the spectral group representation, to the most recent model-based approach by \cite{Jacques2014}, only few works proposed a probabilistic model for the whole set of 15449 partial orderings.
Here we will show at what extent PL mixture models are able to provide an in-depth overall account of the underlying group structure.

We fitted Bayesian PL mixtures with $G=1,\dots,12$ components to the whole APA dataset.
A relatively parsimonious PL mixture is selected by our parametric approach by using the Bayesian selection criteria displayed in Figure~\ref{fig:APAcriteria} (numerical values are reported in Table~\ref{t:BaycritAPA}).
Indeed,
BICM$_1$ and BICM$_2$ agree with BIC in selecting 5 components. However, as suggested in our simulation study, we privilege the use of BPIC$_1$ and DIC$_1$ which both agree in identifying $\hat G=10$ groups. On the other hand, the alternative BNPPLM analysis adopting Dahl's procedure yields a partition of the electorate in 86 distinct clusters.
Prior to illustrating the interpretation of the fitted components, we provide some new insights on model assessment.
For the selected model, $p_{B(1)}=0.471$ and $p_{B(2)}=0.582$ do not highlight overall lack-of-fit; see Table~\ref{t:BaycritAPA}. 
However, the substantial presence of strictly partial orderings suggested a more specific check considering the 
conditional distributions 
of the same univariate and bivariate preference features within each subset $\upi^{-1}_m$ of partial top orderings with the same length $m=1,\dots,4$.
It revealed that the best global model fitted to the whole set of ballots is unsuitable to describe the heterogeneity of these subsets. In fact, the corresponding $\tilde p_{B(1)}<10^{-4}$ and $\tilde p_{B(2)}<10^{-4}$ are well below the conventional 0.05 critical threshold and suggest to implement our PL mixture model separately on each subset, in order to provide a more appropriate account of the heterogeneity in the APA election data. We will then compare these results with those corresponding to our initial PL mixture analysis on the entire dataset. Thus, we estimated the Bayesian PL mixture separately on top-1, top-2, top-3 and top-4 (full) orderings.
Notice that on top-1 orderings only the PL mixture with $G=1$ can be fully identified,
since they correspond to ordinary multinomial data on $K$ categories. Selection of the optimal number of components is displayed in Figure~\ref{fig:APAcriteria} (numerical values are reported in Table~\ref{t:BaycritAPA2}).
Indeed, if we analyze separately each subset and comment overall on all the resulting subgroups, we get a total of $(1+2+3+7)=13$ clusters. We believe that these clusters provide a more appropriate representation of the heterogeneity in the APA election data. Unlike the overall model fitted to all the available ballots, for each model we get satisfactory fitting diagnostics (Table~\ref{t:BaycritAPA2}). 
Now, let us focus on the support parameter estimates of the different components fitted to each subset (Figure~\ref{fig:separate}). We notice that
all the components exhibit distinctive modal orderings, apart from one component which shares the same modal pattern 
(C,A,B,E,D). Only three of them are recovered in the groups fitted to the whole dataset (Figure~\ref{fig:global}). Moreover, in none of the modal orderings of the components fitted with the separate analyses Candidate B is ranked first. Instead, in two out of ten components of the global model (corresponding to a total weight 0.16) Candidate B occupies the first position of the corresponding modal ordering with a relatively large estimated support parameter.
This is, to a certain extent, surprising since Candidate B is that
less frequently ranked first in all the subsets; see Figure~\ref{f:freqmissAPA} (right) and the corresponding Table \ref{t:firstchoice}.
Another interesting evidence from the separate analysis is that almost all components have Candidate C, D, or E in the first position of the modal orderings. The only exception, which provides maximum support to Candidate A, is found in a component fitted to the subset of full orderings. Such exceptional component, with estimated weight $\hat \omega_1=0.05$ in that subset, amounts to 1.9\% of the entire dataset. Additionally, if we aggregate the relative weights of all components resulting from the separate analysis which have Candidate C (the winner of the election) in the first position of the modal ordering, we get a total weight of 0.556. The analogue computation on the global mixture returns a total weight of 0.30. 
Notice, however, that both analyses provide a similar posterior mean vector of the support parameters, equal to $\hat{\underline{p}}=(0.189,0.148,0.259,0.208,0.196)$ in the global analysis and $\hat{\underline{p}}=(0.192,0.148,0.257, 0.206, 0.197)$ with the aggregation of the separate mixtures. 
Finally, by looking specifically at the results of the analysis on the 5738 top-4 (full) orderings, we can compare our findings with those of previous analyses. We found a larger number of components than \cite{Diaconis-Rep} and \cite{Stern1993}, both of whom identified three clusters of voters, whereas \cite{Jacques2014} reported the lowest BIC for ten components, although with a similar BIC corresponding to four components. Indeed, some vectors of support parameters characterizing our group structure well compare with the findings in \cite{Stern1993}, especially those for which Candidate C is in the first position of the modal ordering. Overall, our findings allow for a characterization of the groups in terms of a more marked  preference for one or two candidates.

%

\section{Concluding remarks and future work}
\label{s:concl}

We have investigated a Bayesian finite PL mixture for dealing with
heterogeneous partially ranked data
and described efficient algorithms to conduct posterior inference.
Our proposal contemplates a 
data augmentation step with the latent group structure
and allows for model-based classification of partial top orderings.
It can be seen as a direct
extension to the finite mixture context 
of the
basic Bayesian PL
introduced
by \cite{Caron:Doucet}, 
aimed at identifying and characterizing possible groups of rankers with similar preferences/attitudes.
On the other hand, 
it can be 
regarded as 
a Bayesian generalization 
of the PL mixture developed by~\cite{Gormley:Murphy-Royal},
whose frequentist approach can be recovered
as a by-product of the noninformative analysis.
An important advantage over the MLE perspective lies in the possibility to straightforwardly address estimation uncertainty, without relying on large sample approximations and additional computational burden.

We have investigated the effectiveness of our estimation algorithms 
in a simulation study with multiple heterogeneity scenarios.
In particular, we focused
on the ability 
to recover the actual number of clusters of the generative mixture configuration. Our Bayesian parametric proposal provided a quite satisfactory performance, even when compared with the frequentist approach as well as with the Bayesian nonparametric alternative offered by the BNPPLM 
in \cite{Caron:Teh2014}.
Our simulations
highlighted sometimes remarkable divergences in the final determination of the number of clusters, possibly due to the theoretically different notion of ``group'' behind the two Bayesian models. 
The analysis of two real experiments provided further evidence on the usefulness of our parametric model.
For the CARCONF data, the existence of a heterogeneous pattern of preferences
emerged neither from the BNPPLM nor from the frequentist approach, whereas our proposal identified a 2-component PL mixture 
with two meaningful differential profiles.
In general, estimating a smaller number of groups means that some preference patterns would not be recognized, leading to a less informative picture of the underlying preference system.
On the other hand, the nonparametric method could prove itself more flexible to recover possible departures from the reference parametric ranking distribution by fitting a higher number of minor clusters to the sample.
Summing up, both simulations and real dataset analyses highlighted that
our Bayesian finite PL mixture and the BNPPLM
can lead to substantially different conclusions and, sometimes, our proposal could be preferred. This happens despite the fact that
the nonparametric BNPPLM method could be regarded somehow as a generalization of the Bayesian finite PL mixture.
Additionally, this work provided some incremental findings on the performance of many alternative Bayesian selection criteria. Our investigation suggests, besides the most frequently adopted DIC$_1$, the use of 
BPIC$_1$. 
Also BIC performed well for smaller values of $G^*$. However, for larger values of $G^*$ we confirm
BIC's tendency to underestimate the true number of groups,
as also pointed out in 
other
mixture settings; see for example~\cite{Celeux1996,Lukovciene}
and~\cite{Bulteel}. 
In line with this evidence, 
under Scenario 4 no overestimation is present with BIC; on the other hand, BIC leads to underestimate the true number $G^*$ of components for at least 30\% of the simulated datasets in all the three censoring settings.
Indeed, one could argue that, as a function of the sample size, the penalty term of BIC does not account for the varying rate of truncation, leading to a too severe penalization and, hence, to the selection of more parsimonious models. Conversely, with DIC$_1$ and BPIC$_1$ the effective number of parameters depends on the posterior deviance distribution that inherently penalizes for the increasing parameterization and the higher censoring rate. For this reason, the two Bayesian criteria could be expected to return a more adaptive and suitable estimation of model complexity.
Certainly, a more theoretical advancement is needed before a clear-cut conclusion on the most suitable criterion to adopt in the finite mixture framework, where regularity conditions facilitating the derivation of asymptotic results do not hold. Indeed, apart from few recent attempts \citep{Miller2013,Miller2014}, in the nonparametric setting the asymptotic behavior is even less explored and understood. 

We also made use of diagnostic devices to evaluate the fitting of our proposal 
via a posterior predictive check. 
Despite its practical relevance, the fitting performance
is often neglected by practitioners, especially in the frequentist analysis of ranking data.
Unlike previous applications in the partial ranking literature, we have also applied discrepancy measures accounting for the conditional distributions, given  the number of ranked items. These allow us to gain a more in-depth understanding of the adequacy of the PL parametric assumption in the whole dataset.

A possible future development
could be the Bayesian estimation of the mixture of Extended PL recently introduced by \cite{Mollica:Tardella}. One can extend model flexibility by exploiting the
additional 
\textit{reference order} parameter, 
representing the rank attribution order followed by the ranker to sequentially carry out his comparative judgment on the available items. 
Another interesting extension
could be the introduction of extra information provided by
individual and/or item-specific covariates.
As revealed by previous applications
\citep{Gormley:Murphy-AnnalsApplied,Gormley:Murphy2010},
explanatory variables may fruitfully contribute to characterize 
choice patterns and support
decisions for better capturing specific preference profiles or segments.
Finally, 
the lack-of-fit 
due to the differential preference patterns underlying the different subsets of rankers who provide the same number of partially ranked items highlights the need for a more comprehensive model accounting for this type of observed heterogeneity.
\vspace{\fill}\pagebreak




%




\end{document}


%
%
%
%
%
%
%
%

\setcounter{page}{1}
\vspace*{2\baselineskip}

\RepeatTitle{Bayesian Mixture of Plackett-Luce models
for partially ranked data}\vskip3pt

\author{Cristina Mollica and Luca Tardella}

\affil{Dipartimento di Scienze statistiche, Sapienza Universit\`a di Roma, Rome, Italy} 
\thanks{The final publication is available at Springer via \url{http://dx.doi.org/10.1007/s11336-016-9530-0}}

\renewcommand\thefigure{SM-\arabic{figure}}    
\renewcommand\thetable{SM-\arabic{table}}    


\section{Supplementary Material}
\setcounter{figure}{0}




\subsection{Implementation details for model assessment: posterior predictive checks $p_{B(1)}$ and $p_{B(2)}$}

We remind that the posterior predictive $p$ value 
represents the posterior probability that 
a parameter-dependent discrepancy measure 
$X^2_{(d)}(\upi^{-1}_\text{obs};\theta)$, comparing 
actually observed frequencies and 
expected frequencies under the assumed model $H$,
does not exceed the same discrepancy measure 
$X^2_{(d)}(\upi^{-1}_\text{rep};\theta)$ 
evaluated on a replicated data set drawn from the same model,
that is, 
\begin{equation*}
\label{supp:e:postp}
p_{B(d)}=\PP\left(X_{(d)}^2(\upi_{\text{rep}}^{-1};\theta)\geq X_{(d)}^2(\upi^{-1}_\text{obs};\theta)\biggr\vert\upi^{-1}_\text{obs},H\right).
\end{equation*}
The value $p_{B(d)}$ can be easily approximated by using the posterior
simulations of the parameter vector $\theta$ and augmenting them with 
the corresponding draws of replicated data $\upi_{\text{rep}}^{-1}$.
For our first discrepancy measure $X^2_{(1)}(\upi^{-1};\theta) = \sum_{i=1}^K \frac{\left(r_i(\upi^{-1}) - r_i^*(\theta)\right)^2}
{r_i^*(\theta)}$,
the theoretical frequencies expected under PL mixture model with parameter
$\theta=(\underline{p},\uomega)$ depend on the marginal overall
support parameters 
$p_i=\sum_{g=1}^G\omega_gp_{gi}$ (for $i=1,\dots,K$) and are easily determined as follows:
$$r_i^*(\theta)= N p_i.$$
For the discrepancy measure $X^2_{(2)}(\upi^{-1};\theta)=\sum_{i<i'}\frac{(\tau_{ii'}(\upi^{-1})-\tau^*_{ii'}(\theta))^2}{\tau^*_{ii'}(\theta)}$, 
one can derive 
the expected paired comparison frequencies under PL mixture model
as follows:
$$\tau^*_{ii'}(\theta)=
T_{ii'}\frac{p_i}{p_i+p_{i'}},$$
where 
$T_{ii'}=\tau_{ii'}+\tau_{i'i}$ indicates the total number of pairwise comparisons between item $i$ and $i'$.

\subsection{Implementation details for model assessment: posterior predictive checks $\tilde{p}_{B(1)}$ and $\tilde{p}_{B(2)}$}

Let $m=1,\dots,K-1$ be the generic number of ranked items in a partial ordering of $K$ items.
We denote with $\upi^{-1}_{m}=\{\pi^{-1}_{s}:n_s=m\}$ the subsample of $N_m$ top-$m$ orderings ($\sum_{m=1}^{K-1}N_m=N$). 
In order to assess the model adequacy regarding the homogeneity assumption
on the conditional distributions given the same number $m$ of ranked
items, we define the discrepancy 
between each conditional distribution and the marginal distribution 
of the most-liked item by using the conditional frequencies $r_{i,m}$ as follows:
$$\tilde X^2_{(1)}(\upi^{-1};\theta)=\sum_{m=1}^{K-1}\sum_{i=1}^K\frac{(r_{i,m}-r^*_{i,m}(\theta))^2}{r^*_{i,m}(\theta)},$$ 
where $r_{i,m}=r_i(\upi^{-1}_m)$ and $r^*_{i,m}(\theta)=N_mp_i$.
Similarly, when we aim at assessing homogeneity of the conditional pairwise comparison frequencies, we define 
$$\tilde X^2_{(2)}(\upi^{-1};\theta)= \sum_{m=1}^{K-1}\sum_{i<i'}\frac{(\tau_{ii',m}-\tau^*_{ii',m}(\theta))^2}{\tau^*_{ii',m}(\theta)},$$ 
where $\tau_{ii',m}=\tau_{ii'}(\upi^{-1}_m)$ and $\tau^*_{ii',m}(\theta)=T_{ii',m}\frac{p_i}{p_i+p_{i'}}$ with $T_{ii',m}=\tau_{ii',m}+\tau_{i'i,m}$.
The computation of $\tilde{p}_{B(1)}$ and $\tilde{p}_{B(2)}$ follows the general formula \eqref{e:postp} in the main paper by replacing the desired discrepancy measure.

\section{Supplemental tables and figures}

\vspace*{-15in}

\begin{table}[h!]
\caption{\scriptsize {\em Simulation study (Censoring setting A)} Distribution (\%) of the optimal number $\hat G$ of components identified, respectively, by the Bayesian PL mixture analysis via alternative model selection criteria and by the BNPPLM analysis via Dahl's procedure. In the simulation study, 100 data sets with 1000 partial orderings of 6 items were generated from each PL mixture scenario with alternative true number $G^*$ of components. Best agreement rates under each simulation scenario are highlighted in bold.}
\scriptsize 
\label{t:Distr1}
\centering
\begin{tabular}
 {lcccccccc}
  \multicolumn{9}{c}{$G^*=1$} \\
 \hline 
$\hat G$ & DIC$_1$ & DIC$_2$ & BPIC$_1$ & BPIC$_2$ & BICM$_1$ & BICM$_2$ & BIC & BNPPLM \\
\hline
  1 & 99 & 98 & \bf{100} & \bf{100} & \bf{100} & 99 & \bf{100} & \bf{100} \\
  2 & 1 & 2 & 0 & 0 & 0 & 1 & 0 & 0\\
  3 & 0 & 0 & 0 & 0 & 0 & 0 & 0 & 0\\ 
  4 & 0 & 0 & 0 & 0 & 0 & 0 & 0 & 0\\ 
  5 & 0 & 0 & 0 & 0 & 0 & 0 & 0 & 0\\ 
  6 & 0 & 0 & 0 & 0 & 0 & 0 & 0 & 0\\ 
  7 & 0 & 0 & 0 & 0 & 0 & 0 & 0 & 0\\ 
\multicolumn{9}{c}{$G^*=2$} \\
 \hline 
$\hat G$ & DIC$_1$ & DIC$_2$ & BPIC$_1$ & BPIC$_2$ & BICM$_1$ & BICM$_2$ & BIC & BNPPLM  \\
\hline
1 & 2 & 2 & 2 & 2 & 6 & 6 & 3 & 4\\
2 & 96 & 93 & \bf{98} & 95 & 93 & 89 & 97 & 91\\
3 & 2 & 3 & 0 & 3 & 1 & 3 & 0 & 4\\
4 & 0 & 1 & 0 & 0 & 0 & 2 & 0 & 0\\
5 & 0 & 0 & 0 & 0 & 0 & 0 & 0 & 1\\
6 & 0 & 0 & 0 & 0 & 0 & 0 & 0 & 0\\
7 & 0 & 0 & 0 & 0 & 0 & 0 & 0 & 0\\ 
\multicolumn{9}{c}{$G^*=3$} \\
\hline 
$\hat G$ & DIC$_1$ & DIC$_2$ & BPIC$_1$ & BPIC$_2$ & BICM$_1$ & BICM$_2$ & BIC & BNPPLM  \\
\hline
1 & 0 & 0 & 0 & 0 & 0 & 0 & 0 & 0\\ 
2 & 4 & 4 & 5 & 6 & 8 & 8 & 7 & 6\\ 
3 & 91 & 89 & \bf{94} & 92 & 91 & 88 & 93 & 88\\ 
4 & 5 & 5 & 1 & 2 & 1 & 3 & 0 & 5\\ 
5 & 0 & 1 & 0 & 0 & 0 & 1 & 0 & 1\\ 
6 & 0 & 0 & 0 & 0 & 0 & 0 & 0 & 0 \\ 
7 & 0 & 1 & 0 & 0 & 0 & 0 & 0 & 0 \\ 
    \multicolumn{9}{c}{$G^*=4$} \\
\hline 
$\hat G$ & DIC$_1$ & DIC$_2$ & BPIC$_1$ & BPIC$_2$ & BICM$_1$ & BICM$_2$ & BIC & BNPPLM  \\
\hline
1 & 0 & 0 & 1 & 1 & 1 & 1 & 1 & 1\\ 
2 & 0 & 0 & 0 & 0 & 3 & 3 & 3 & 1\\ 
3 & 18 & 14 & 22 & 22 & 30 & 30 & 30 & 25\\ 
4 & \bf{81} & 67 & 77 & 70 & 65 & 60 & 66 & 50\\ 
5 & 1 & 10 & 0 & 3 & 1 & 6 & 0 & 19\\ 
6 & 0 & 7 & 0 & 4 & 0 & 0 & 0 & 3\\ 
7 & 0 & 2 & 0 & 0 & 0 & 0 & 0 & 1\\ 
 \hline
\end{tabular}
\end{table}

\vspace*{-15in}

\begin{table}[h!]
\caption{\scriptsize {\em Simulation study (Censoring setting B)} Distribution (\%) of the optimal number $\hat G$ of components identified, respectively, by the Bayesian PL mixture analysis via alternative model selection criteria and by the BNPPLM analysis via Dahl's procedure. In the simulation study, 100 data sets with 1000 partial orderings of 6 items were generated from each PL mixture scenario with alternative true number $G^*$ of components. Best agreement rates under each simulation scenario are highlighted in bold.}
\scriptsize 
\label{t:Distr2}
\centering
\begin{tabular}
 {lcccccccc}
  \multicolumn{9}{c}{$G^*=1$} \\
 \hline 
$\hat G$ & DIC$_1$ & DIC$_2$ & BPIC$_1$ & BPIC$_2$ & BICM$_1$ & BICM$_2$ & BIC & BNPPLM \\
\hline
1 & 99 & \bf{100} & \bf{100} & \bf{100} & \bf{100} & 98 & \bf{100} & \bf{100} \\ 
2 & 1 & 0 & 0 & 0 & 0 & 2 & 0 & 0 \\ 
3 & 0 & 0 & 0 & 0 & 0 & 0 & 0 & 0 \\ 
4 & 0 & 0 & 0 & 0 & 0 & 0 & 0 & 0 \\ 
5 & 0 & 0 & 0 & 0 & 0 & 0 & 0 & 0 \\ 
6 & 0 & 0 & 0 & 0 & 0 & 0 & 0 & 0 \\ 
7 & 0 & 0 & 0 & 0 & 0 & 0 & 0 & 0 \\ 
\multicolumn{9}{c}{$G^*=2$} \\
 \hline 
$\hat G$ & DIC$_1$ & DIC$_2$ & BPIC$_1$ & BPIC$_2$ & BICM$_1$ & BICM$_2$ & BIC & BNPPLM  \\
\hline
1 & 2 & 2 & 3 & 2 & 5 & 5  & 5 & 4\\ 
2 & 97 & 93 & 97 & \bf{98} & 95 & 89 & 95 & 89\\ 
3 & 1 & 3 & 0 & 0 & 0 & 5 & 0 & 6\\ 
4 & 0 & 2 & 0 & 0 & 0 & 1 & 0 & 1\\ 
5 & 0 & 0 & 0 & 0 & 0 & 0 & 0 & 0 \\ 
6 & 0 & 0 & 0 & 0 & 0 & 0 & 0 & 0 \\ 
7 & 0 & 0 & 0 & 0 & 0 & 0 & 0 & 0 \\ 
\multicolumn{9}{c}{$G^*=3$} \\
\hline 
$\hat G$ & DIC$_1$ & DIC$_2$ & BPIC$_1$ & BPIC$_2$ & BICM$_1$ & BICM$_2$ & BIC & BNPPLM  \\
\hline
1 & 0 & 0 & 0 & 0 & 0 & 0 & 0 & 0 \\ 
2 & 5 & 5 & 5 & 6 & 10 & 11 & 8 & 7\\ 
3 & 92 & 88 & \bf{94} & 92 & 90 & 82 & 92 & 79\\ 
4 & 3 & 5 & 1 & 2 & 0 & 7 & 0 & 11\\ 
5 & 0 & 2 & 0 & 0 & 0 & 0 & 0 & 2\\ 
6 & 0 & 0 & 0 & 0 & 0 & 0 & 0 & 0\\ 
7 & 0 & 0 & 0 & 0 & 0 & 0 & 0 & 1\\ 
    \multicolumn{9}{c}{$G^*=4$} \\
\hline 
$\hat G$ & DIC$_1$ & DIC$_2$ & BPIC$_1$ & BPIC$_2$ & BICM$_1$ & BICM$_2$ & BIC & BNPPLM  \\
\hline
1 & 1 & 1 & 1 & 1 & 1 & 1 & 1 & 1 \\ 
2 & 0 & 0 & 0 & 1 & 6 & 6 & 5 & 2\\ 
3 & 21 & 21 & 30 & 30 & 39 & 39 & 37 & 19\\ 
4 & \bf{76} & 61 & 69 & 65 & 53 & 48 & 57 & 50\\ 
5 & 2 & 9 & 0 & 3 & 1 & 6 & 0 & 20\\ 
6 & 0 & 7 & 0 & 0 & 0 & 0 & 0 & 5\\ 
7 & 0 & 1 & 0 & 0 & 0 & 0 & 0 & 2\\ 
8 & 0 & 0 & 0 & 0 & 0 & 0 & 0 & 1\\ 
 \hline
\end{tabular}
\end{table}

\vspace*{-15in}

\begin{table}[h!]
\caption{\scriptsize {\em Simulation study (Censoring setting C)} Distribution (\%) of the optimal number $\hat G$ of components identified, respectively, by the Bayesian PL mixture analysis via alternative model selection criteria and by the BNPPLM analysis via Dahl's procedure. In the simulation study, 100 data sets with 1000 partial orderings of 6 items were generated from each PL mixture scenario with alternative true number $G^*$ of components. Best agreement rates under each simulation scenario are highlighted in bold.}
\scriptsize 
\label{t:Distr3}
\centering
\begin{tabular}
 {lcccccccc}
  \multicolumn{9}{c}{$G^*=1$} \\
 \hline 
$\hat G$ & DIC$_1$ & DIC$_2$ & BPIC$_1$ & BPIC$_2$ & BICM$_1$ & BICM$_2$ & BIC & BNPPLM \\
  \hline
1 & 97 & 98 & 98 & \bf{100} & \bf{100} & \bf{100} & \bf{100} & \bf{100}\\ 
2 & 3 & 1 & 2 & 0 & 0 & 0 & 0 & 0 \\ 
3 & 0 & 0 & 0 & 0 & 0 & 0 & 0 & 0 \\ 
4 & 0 & 0 & 0 & 0 & 0 & 0 & 0 & 0 \\ 
5 & 0 & 0 & 0 & 0 & 0 & 0 & 0 & 0 \\ 
6 & 0 & 1 & 0 & 0 & 0 & 0 & 0 & 0 \\ 
7 & 0 & 0 & 0 & 0 & 0 & 0 & 0 & 0 \\ 
\multicolumn{9}{c}{$G^*=2$} \\
 \hline 
$\hat G$ & DIC$_1$ & DIC$_2$ & BPIC$_1$ & BPIC$_2$ & BICM$_1$ & BICM$_2$ & BIC & BNPPLM  \\
\hline
1 & 2 & 2 & 3 & 3 & 8 & 8 & 5 & 5 \\ 
2 & \bf{97} & 91 & \bf{97} & 95 & 91 & 87 & 95 & 90 \\ 
3 & 1 & 6 & 0 & 2 & 1 & 5 & 0 & 5 \\ 
4 & 0 & 1 & 0 & 0 & 0 & 0 & 0 & 0 \\ 
5 & 0 & 0 & 0 & 0 & 0 & 0 & 0 & 0 \\ 
6 & 0 & 0 & 0 & 0 & 0 & 0 & 0 & 0 \\ 
7 & 0 & 0 & 0 & 0 & 0 & 0 & 0 & 0 \\ 
\multicolumn{9}{c}{$G^*=3$} \\
\hline 
$\hat G$ & DIC$_1$ & DIC$_2$ & BPIC$_1$ & BPIC$_2$ & BICM$_1$ & BICM$_2$ & BIC & BNPPLM  \\
\hline
1 & 0 & 0 & 0 & 0 & 0 & 0 & 0 & 0 \\ 
2 & 4 & 5 & 6 & 6 & 12 & 11 & 8 & 8 \\ 
3 & \bf{95} & 89 & 94 & 92 & 88 & 82 & 92 & 63 \\ 
4 & 1 & 5 & 0 & 2 & 0 & 5 & 0 & 22 \\ 
5 & 0 & 1 & 0 & 0 & 0 & 1 & 0 & 6 \\ 
6 & 0 & 0 & 0 & 0 & 0 & 0 & 0 & 1 \\ 
7 & 0 & 0 & 0 & 0 & 0 & 1 & 0 & 0 \\ 
    \multicolumn{9}{c}{$G^*=4$} \\
\hline 
$\hat G$ & DIC$_1$ & DIC$_2$ & BPIC$_1$ & BPIC$_2$ & BICM$_1$ & BICM$_2$ & BIC & BNPPLM  \\
\hline
1 & 0 & 0 & 1 & 1 & 2 & 2  & 1 & 1 \\ 
2 & 1 & 1 & 0 & 1 & 9 & 9 & 6 & 2\\ 
3 & 25 & 23 & 32 & 30 & 41 & 39 & 38 & 22\\ 
4 & \bf{72} & 62 & 67 & 64 & 48 & 46 & 55 & 37\\ 
5 & 2 & 7 & 0 & 3 & 0 & 1 & 0 & 16\\ 
6 & 0 & 6 & 0 & 1 & 0 & 0 & 0 & 13\\ 
7 & 0 & 1 & 0 & 0 & 0 & 3 & 0 & 6\\ 
8 & 0 & 0 & 0 & 0 & 0 & 0 & 0 & 3\\ 
 \hline
\end{tabular}
\end{table}

\clearpage

\begin{figure}[h]
\centering
\subfloat{ 
\includegraphics[scale=.42]{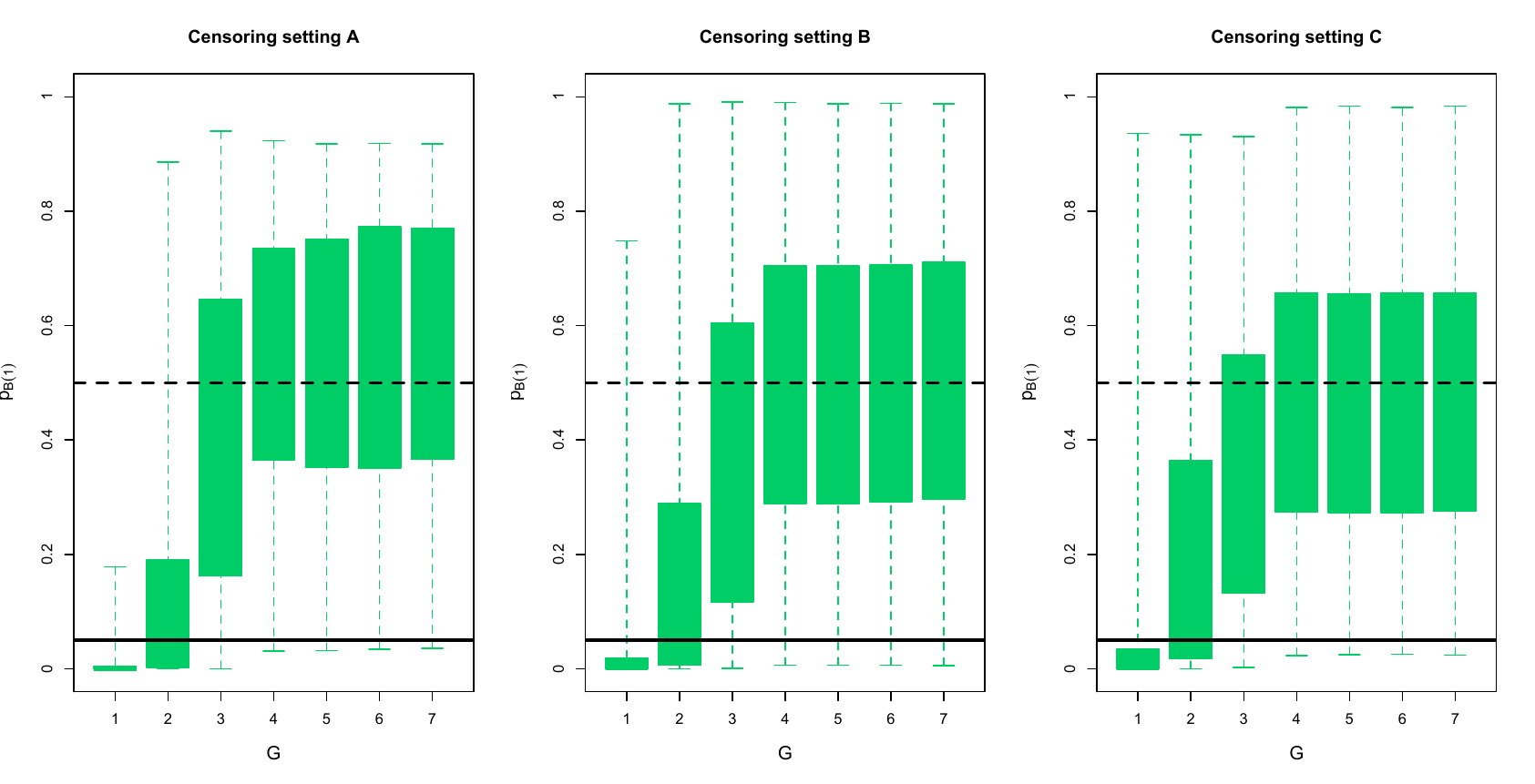}
\label{fig:Diagnostic1}}\\
\subfloat{ 
\includegraphics[scale=.42]{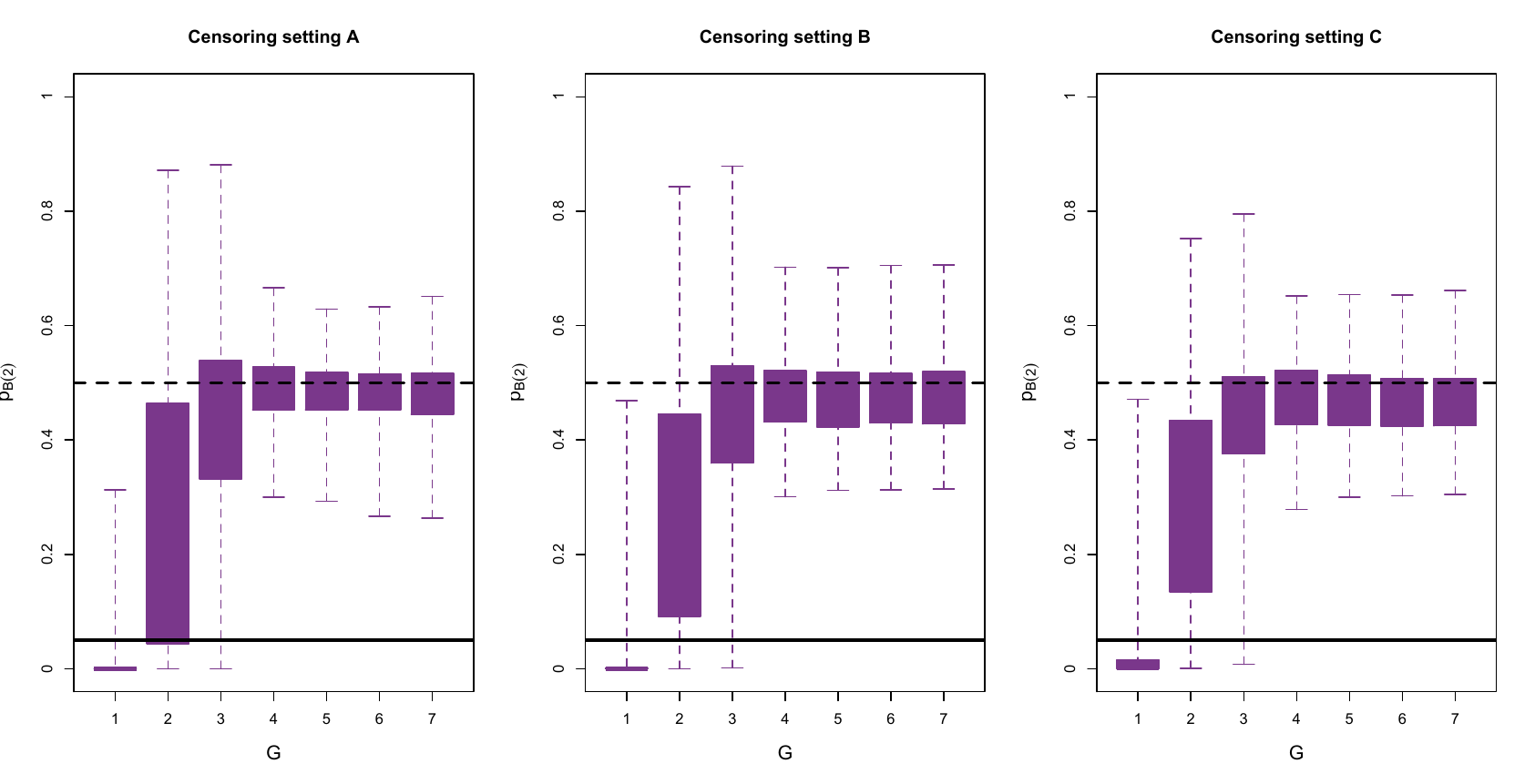}
\label{fig:Diagnostic2}}
\caption{{\em Simulation study (Scenario 4)} Model assessment criteria with a varying number $G$ of fitted components for the Bayesian PL mixtures fitted to the 100 data sets simulated from the population scenario with $G^*=4$ groups. The solid and dashed lines represent, respectively, the critical threshold 0.05 and the reference value 0.5 expected under correct model specification.}
\label{fig:Diagnostic}
\end{figure}
%

 %
\begin{table}[t]
\caption{{\em CARCONF data (435 full and partial orderings)} Model selection criteria and posterior
  predictive $p$ values for the Bayesian PL mixtures with a
  varying number $G$ of components. 
For each selection criterion, 
the optimal choice of the number of components corresponds 
to the minimum value (\textit{in bold}). 
}
\scriptsize
\renewcommand{\arraystretch}{1} 
\label{t:Baycrit3}
\centering
\begin{tabular}
 {lccccccccc} 
$G$ & DIC$_1$ & DIC$_2$ & BPIC$_1$ & BPIC$_2$ & BICM$_1$ & BICM$_2$ & BIC & $p_{B(1)}$ & $p_{B(2)}$ \\
\hline
1 & 5288.34 & 5288.29 & 5293.32 & 5293.24 & \bf{5308.44} & \bf{5308.39} & \bf{5308.74} & 0.000 & 0.247 \\ 
2 & \bf{5268.73} & \bf{5268.90} & \bf{5280.15} & \bf{5280.48} & 5316.09 & 5316.25 & 5312.73 & 0.079 & 0.505 \\ 
3 & 5278.45 & 5273.38 & 5301.99 & 5291.84 & 5348.62 & 5343.55 & 5334.66 & 0.092 & 0.515 \\ 
4 & 5289.34 & 5272.67 & 5324.82 & 5291.47 & 5349.29 & 5332.61 & 5358.12 & 0.103 & 0.508 \\ 
5 & 5295.06 & 5273.46 & 5336.93 & 5293.75 & 5356.14 & 5334.55 & 5387.49 & 0.107 & 0.516 \\ 
6 & 5305.01 & 5274.43 & 5357.28 & 5296.12 & 5362.83 & 5332.25 & 5413.11 & 0.122 & 0.518 \\ 
\hline
\end{tabular}
\end{table}
\normalsize

\begin{table}[h!]
\scriptsize 
\caption{{\em CARCONF data (435 full and partial orderings)} Posterior means of the parameters and
  component-specific modal orderings of the optimal Bayesian
  2-component PL mixture. 
Posterior standard deviations are shown in parentheses.}
\centering
\begin{tabular}
 {crlcrlrlrlrlrlrl}
 $g$ & $\hat\omega_g$ & & $\hat\sigma_g^{-1}$ & $\hat p_{g1}$ & & $\hat p_{g2}$ & & $\hat p_{g3}$ & & $\hat p_{g4}$ & & $\hat p_{g5}$ & & $\hat p_{g6}$ & \\
\hline
1 & .713 & (.10) & (2,6,4,3,1,5)  & .079 & (.02) & .263 & (.02) & .185 & (.02) & .191 & (.01)  & .071  & (.01) & .211 & (.02) \\
2 & .287 & (.10) & (1,3,4,2,6,5)  & .436 & (.13) & .124 & (.04) & .157 & (.05) & .138 & (.03)  & .043  & (.02) & .101 & (.03)  \\
\end{tabular}
\label{t:carest}
\end{table}

 \scriptsize
  \begin{table}[h!]
\caption{{\em APA election data (15449 full and partial orderings)} Percentage of voters who assign position $t$ to Candidate $i$ (\textit{upper panel}) and average rank vector (\textit{lower panel}).}
\small 
\centering
\begin{tabular}{rrrrrr}
& \multicolumn{5}{c}{Candidate} \\ 
\cline{2-6}  
Rank & A & B & C & D & E \\ 
 \hline
1 & 18.8 & 14.8 & 26.0 & 21.0 & 19.4 \\ 
2 & 27.7 & 17.7 & 16.9 & 16.9 & 20.7 \\ 
3 & 23.6 & 24.1 & 14.0 & 18.6 & 19.7 \\ 
4 & 17.5 & 24.7 & 18.3 & 20.3 & 19.3 \\ 
5 & 14.8 & 18.4 & 23.1 & 23.4 & 20.3 \\ 
\hline
$\bar \pi$ & 2.37 & 2.66 & 2.34 & 2.51 & 2.47 \\
\end{tabular}
\label{t:firstorder}
\end{table}  

\begin{table}[h!]
\small 
\caption{{\em APA election data (15449 full and partial orderings)} Percentage of voters who rank Candidate $i$ in the first position conditionally on the number $m$ of ranked candidates.}
\centering
\begin{tabular}{rrrrrr}
& \multicolumn{5}{c}{Candidate} \\ 
\cline{2-6}  
 $m$ & A & B & C & D & E \\ 
\hline
1 & 17.4 & 17.1 & 23.3 & 22.3 & 19.9 \\ 
2 & 21.6 & 11.8 & 31.9 & 18.8 & 16.0 \\ 
3 & 20.1 & 16.3 & 20.1 & 21.8 & 21.7 \\ 
4 & 18.4 & 13.5 & 28.0 & 20.4 & 19.7 \\ 
\hline
\end{tabular}
\label{t:firstchoice}
\end{table}
\normalsize


\begin{table}[h!]
\scriptsize 
\caption{{\em APA election data (global analysis with 15449 full and partial orderings)} Model selection criteria and posterior
predictive $p$ values for the Bayesian PL mixtures with a
varying number $G$ of components. 
For each selection criterion, 
the optimal choice of the number of components corresponds 
to the minimum value (\textit{in bold}). 
}
\label{t:BaycritAPA}  
\begin{tabular}{lccccccccc}
\multicolumn{10}{c}{Full + Partial orderings} \\
\hline
$G$ & DIC$_1$ & DIC$_2$ & BPIC$_1$ & BPIC$_2$ & BICM$_1$ & BICM$_2$ & BIC & $p_{B(1)}$ & $p_{B(2)}$ \\
\hline
1 & 103204.59 & 103204.65 & 103208.57 & 103208.71 & 103235.66 & 103235.73 & 103235.19 & 0.000 &  0.000 \\ 
2 & 100772.97 & 100772.87 & 100781.63 & 100781.44 & 100838.40 & 100838.30 & 100842.44 & 0.000 & 0.610 \\ 
3 & 100591.84 & 100591.89 & 100603.00 & 100603.10 & 100677.58 & 100677.62 & 100704.56 & 0.004 & 0.493 \\ 
4 & 100436.20 & 100445.60 & 100443.54 & 100462.34 & 100573.58 & 100582.98 & 100604.78 & 0.298 & 0.411 \\ 
5 & 100396.98 & 100401.44 & 100413.46 & 100422.39 & \bf{100561.56} & \bf{100566.03} & \bf{100595.51} & 0.343 & 0.523 \\ 
6 & 100360.59 & 100369.50 & 100377.16 & 100394.99 & 100564.33 & 100573.24 & 100607.17 & 0.375 & 0.503 \\ 
7 & 100336.08 & 100344.30 & 100361.35 & 100377.81 & 100600.44 & 100608.66 & 100613.47 & 0.233 & 0.536 \\ 
8 & 100341.12 & 100347.07 & 100381.82 & 100393.71 & 100703.71 & 100709.65 & 100635.88 & 0.390 & 0.492 \\ 
9 & 100341.55 & 100350.84 & 100390.59 & 100409.18 & 100796.84 & 100806.13 & 100667.85 & 0.426 & 0.531 \\ 
10 & \bf{100317.68} & 100346.31 & \bf{100358.35} & 100415.63 & 100876.22 & 100904.86 & 100708.95 & 0.471 & 0.528 \\ 
11 & 100321.54 & \bf{100314.49} & 100376.02 & \bf{100361.92} & 100677.10 & 100670.05 & 100733.43 & 0.382 & 0.531 \\ 
12 & 100340.38 & 100349.80 & 100408.73 & 100427.57 & 100944.37 & 100953.79 & 100772.76 & 0.440 & 0.521 \\ 
   \hline
\end{tabular}
\end{table}

\begin{table}[h!]
\centering
\begin{minipage}{\textwidth}
\caption{{\em APA election data (separate analysis for subsets of
ballots with the same number of ranked candidates)} Model selection criteria and posterior
  predictive $p$ values for the Bayesian PL mixtures with a
  varying number $G$ of components. 
For each selection criterion, 
the optimal choice of the number of components corresponds 
to the minimum value (\textit{in bold}). 
}
\label{t:BaycritAPA2}  
\end{minipage}
\begin{center}
\scriptsize
\renewcommand{\arraystretch}{0.9}
\begin{tabular}{lccccccccc}
  \multicolumn{10}{c}{Top-2 orderings} \\
\hline
$G$ & DIC$_1$ & DIC$_2$ & BPIC$_1$ & BPIC$_2$ & BICM$_1$ & BICM$_2$ & BIC & $p_{B(1)}$ & $p_{B(2)}$ \\
\hline
1 & 14255.36 & 14255.24 & 14259.33 & 14259.09 & 14277.59 & 14277.47 & 14278.66 & 0.000 & 0.049 \\ 
2 & 13427.79 & 13427.99 & \bf{13436.89} & \bf{13437.29} & \bf{13481.98} & \bf{13482.18} & \bf{13479.88} & 0.295 & 0.502 \\ 
3 & 13427.70 & 13426.86 & 13443.00 & 13441.33 & 13510.92 & 13510.08 & 13506.41 & 0.379 & 0.530 \\ 
4 & 13434.10 & 13427.37 & 13458.77 & 13445.32 & 13531.62 & 13524.90 & 13533.12 & 0.410 & 0.530 \\ 
5 & 13433.24 & 13425.80 & 13458.63 & 13443.74 & 13530.04 & 13522.60 & 13569.87 & 0.443 & 0.531 \\ 
6 & 13431.30 & 13426.03 & 13455.70 & 13445.16 & 13537.16 & 13531.89 & 13608.96 & 0.455 & 0.527 \\ 
7 & 13429.54 & 13425.14 & 13453.09 & 13444.29 & 13536.39 & 13531.99 & 13647.93 &0.463 & 0.522 \\ 
8 & 13429.62 & 13425.23 & 13453.22 & 13444.45 & 13536.84 & 13532.45 & 13686.96 &0.472 & 0.526 \\ 
9 & 13428.02 & \bf{13423.42} & 13450.81 & 13441.60 & 13529.06 & 13524.45 & 13726.03 & 0.478 & 0.525 \\ 
10 & 13427.65 & 13424.41 & 13450.28 & 13443.79 & 13537.01 & 13533.76 & 13765.02 & 0.479 & 0.522 \\ 
11 & 13427.58 & 13423.66 & 13450.16 & 13442.32 & 13532.07 & 13528.15 & 13804.08 & 0.476 & 0.525 \\ 
12 & \bf{13426.88} & 13423.49 & 13449.11 & 13442.32 & 13532.91 & 13529.51 & 13843.13 & 0.478 & 0.513 \\ 
  \multicolumn{10}{c}{Top-3 orderings} \\
\hline
$G$ & DIC$_1$ & DIC$_2$ & BPIC$_1$ & BPIC$_2$ & BICM$_1$ & BICM$_2$ & BIC & $p_{B(1)}$ & $p_{B(2)}$ \\
  \hline
1 & 17147.88 & 17147.84 & 17151.92 & 17151.83 & 17170.41 & 17170.37 & 17170.43 & 0.000 & 0.180 \\ 
2 & 16732.01 & 16733.01 & 16741.62 & 16743.63 & \bf{16793.05} & \bf{16794.05} & \bf{16781.65} & 0.262 & 0.468 \\ 
3 & \bf{16717.06} & 16717.76 & \bf{16731.79} & \bf{16733.19} & 16805.00 & 16805.69 & 16794.74 & 0.278 & 0.529 \\ 
4 & 16717.27 & 16719.43 & 16742.81 & 16747.13 & 16876.03 & 16878.19 & 16811.62 & 0.440 & 0.530 \\ 
5 & 16718.74 & 16720.82 & 16752.55 & 16756.70 & 16923.69 & 16925.77 & 16834.81 & 0.479 & 0.523 \\ 
6 & 16723.28 & 16713.24 & 16762.76 & 16742.70 & 16879.76 & 16869.73 & 16866.26 & 0.483 & 0.511 \\ 
7 & 16725.94 & 16716.31 & 16769.12 & 16749.85 & 16905.96 & 16896.33 & 16899.80 & 0.477 & 0.516 \\ 
8 & 16726.43 & 16715.82 & 16771.67 & 16750.46 & 16911.66 & 16901.06 & 16934.43 & 0.468 & 0.509 \\ 
9 & 16729.16 & 16715.99 & 16776.54 & 16750.20 & 16909.40 & 16896.23 & 16971.16 & 0.497 & 0.509 \\ 
10 & 16735.15 & 16716.82 & 16788.59 & 16751.92 & 16915.26 & 16896.92 & 17003.31 & 0.491 &  0.511 \\ 
11 & 16736.19 & 16718.88 & 16790.71 & 16756.09 & 16929.25 & 16911.94 & 17040.45 & 0.478 & 0.507 \\ 
12 & 16735.28 & \bf{16710.82} & 16790.19 & 16741.29 & 16883.04 & 16858.58 & 17077.01 & 0.497 & 0.505 \\ 
\multicolumn{10}{c}{Top-4 (full) orderings} \\
\hline
$G$ & DIC$_1$ & DIC$_2$ & BPIC$_1$ & BPIC$_2$ & BICM$_1$ & BICM$_2$ & BIC & $p_{B(1)}$ & $p_{B(2)}$ \\
\hline
1 & 54812.60 & 54812.60 & 54816.60 & 54816.61 & 54839.29 & 54839.30 & 54839.21 & 0.000 &  0.000 \\ 
2 & 53696.16 & 53695.65 & 53705.49 & 53704.48 & 53754.42 & 53753.91 & 53755.38 & 0.000 & 0.639 \\ 
3 & 53576.87 & 53574.99 & 53591.48 & 53587.73 & 53659.78 & 53657.91 & 53668.81 & 0.043 & 0.655 \\ 
4 & 53477.70 & 53477.33 & 53494.38 & 53493.63 & 53585.83 & 53585.46 & \bf{53608.79} & 0.179 & 0.478 \\ 
5 & 53458.70 & 53454.30 & 53479.35 & 53470.54 & \bf{53562.39} & \bf{53557.98} & 53625.13 & 0.183 & 0.470 \\ 
6 & 53433.51 & 53439.25 & 53460.29 & 53471.77 & 53655.67 & 53661.42 & 53630.95 & 0.462 & 0.479 \\ 
7 & \bf{53412.08} & 53437.63 & \bf{53445.60} & 53496.70 & 53830.73 & 53856.28 & 53639.31 & 0.503 & 0.512 \\ 
8 & 53420.25 & \bf{53416.43} & 53464.62 & \bf{53456.97} & 53686.23 & 53682.40 & 53669.06 & 0.440 & 0.436 \\ 
9 & 53449.15 & 53499.96 & 53512.83 & 53614.45 & 54261.90 & 54312.71 & 53702.59 & 0.413 & 0.489 \\ 
10 & 53415.29 & 53443.76 & 53466.78 & 53523.72 & 53975.90 & 54004.37 & 53736.39 & 0.481 & 0.553 \\ 
11 & 53437.74 & 53446.27 & 53503.70 & 53520.77 & 53942.07 & 53950.61 & 53773.17 & 0.403 & 0.462 \\ 
12 & 53424.85 & 53438.45 & 53488.02 & 53515.22 & 53949.36 & 53962.97 & 53809.15 & 0.455 & 0.519 \\ 
\hline
\end{tabular}
\end{center}
\end{table}

\begin{figure}[t]
\centering
\subfloat{ 
\includegraphics[scale=.4]{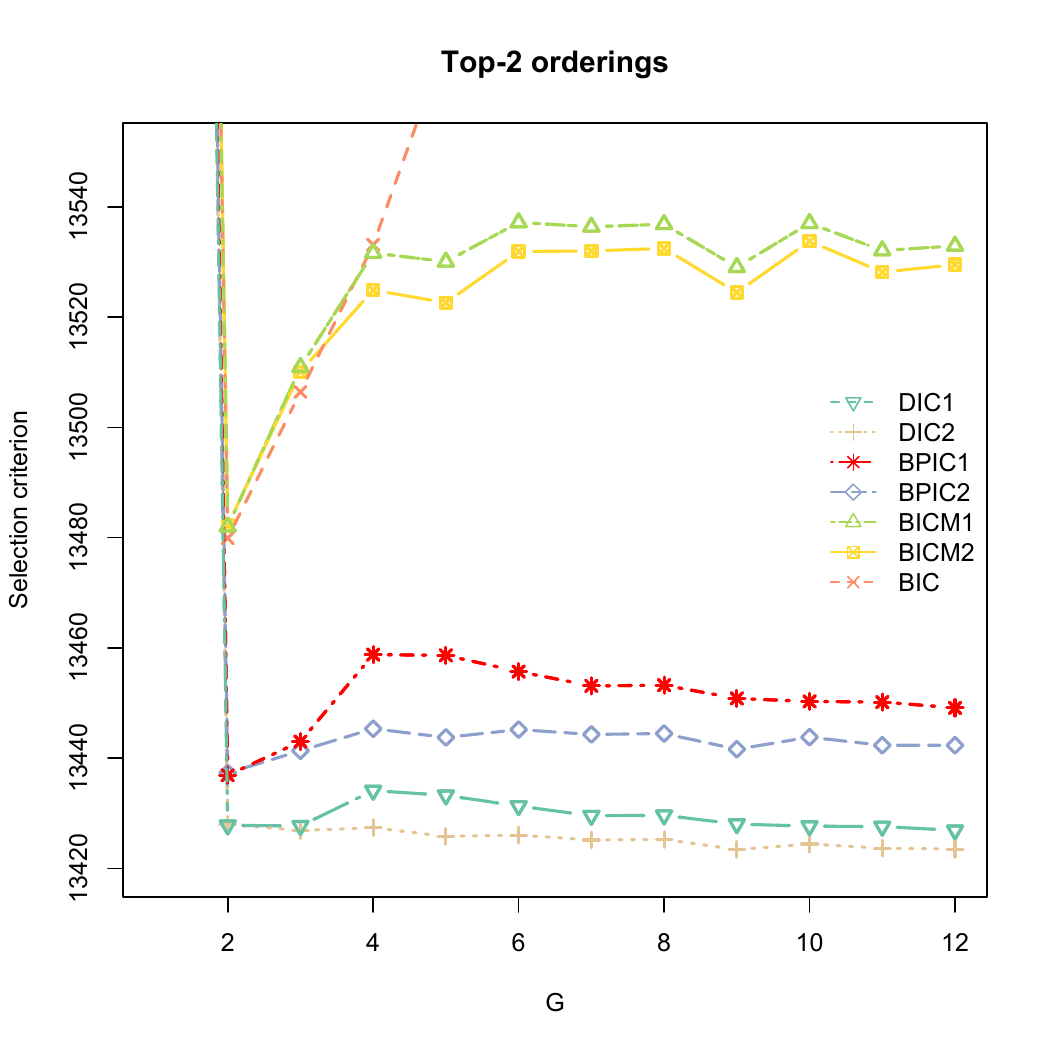}
\label{fig:APAtop2}}
\subfloat{ 
\includegraphics[scale=.4]{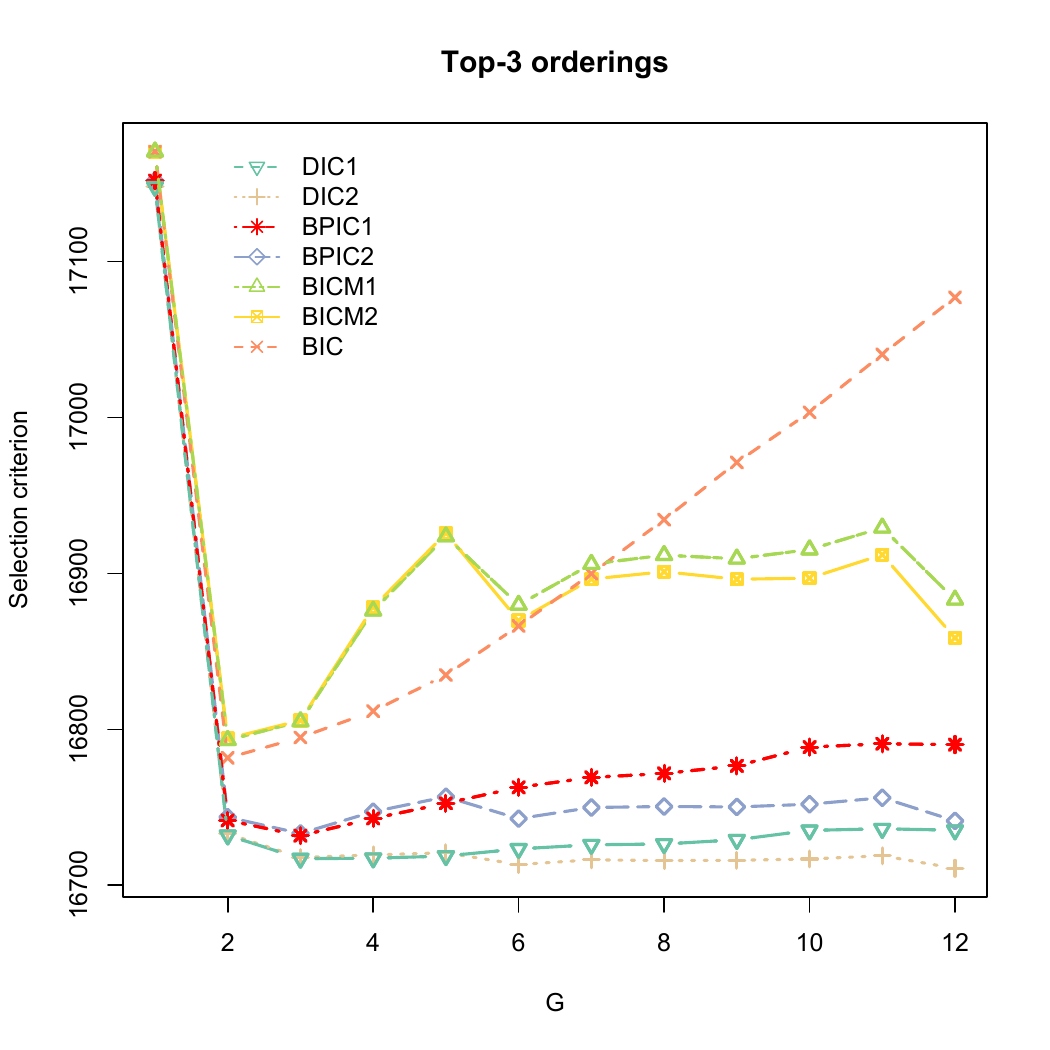}
\label{fig:APAtop3}}\\
\subfloat{ 
\includegraphics[scale=.4]{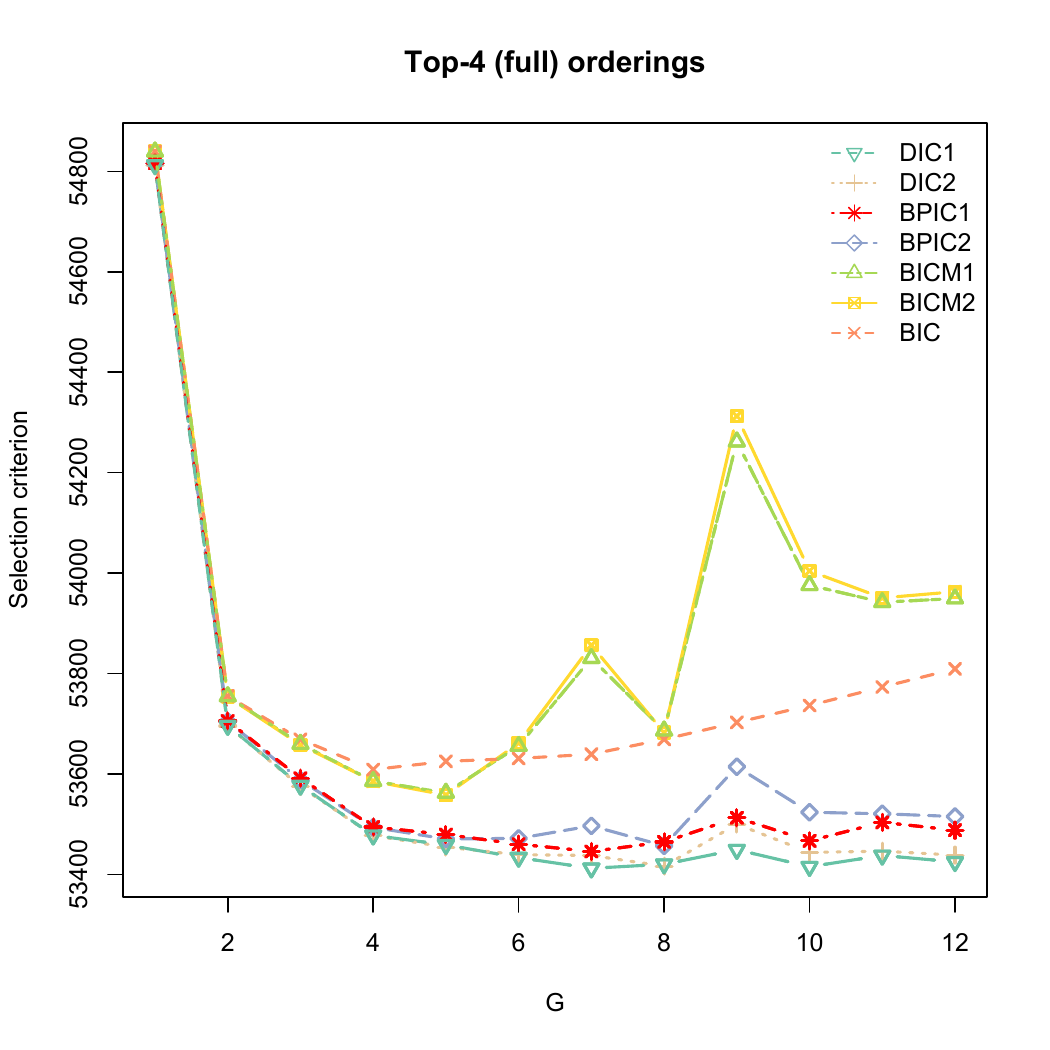}
\label{fig:APAtop4}}
\subfloat{ 
\includegraphics[scale=.4]{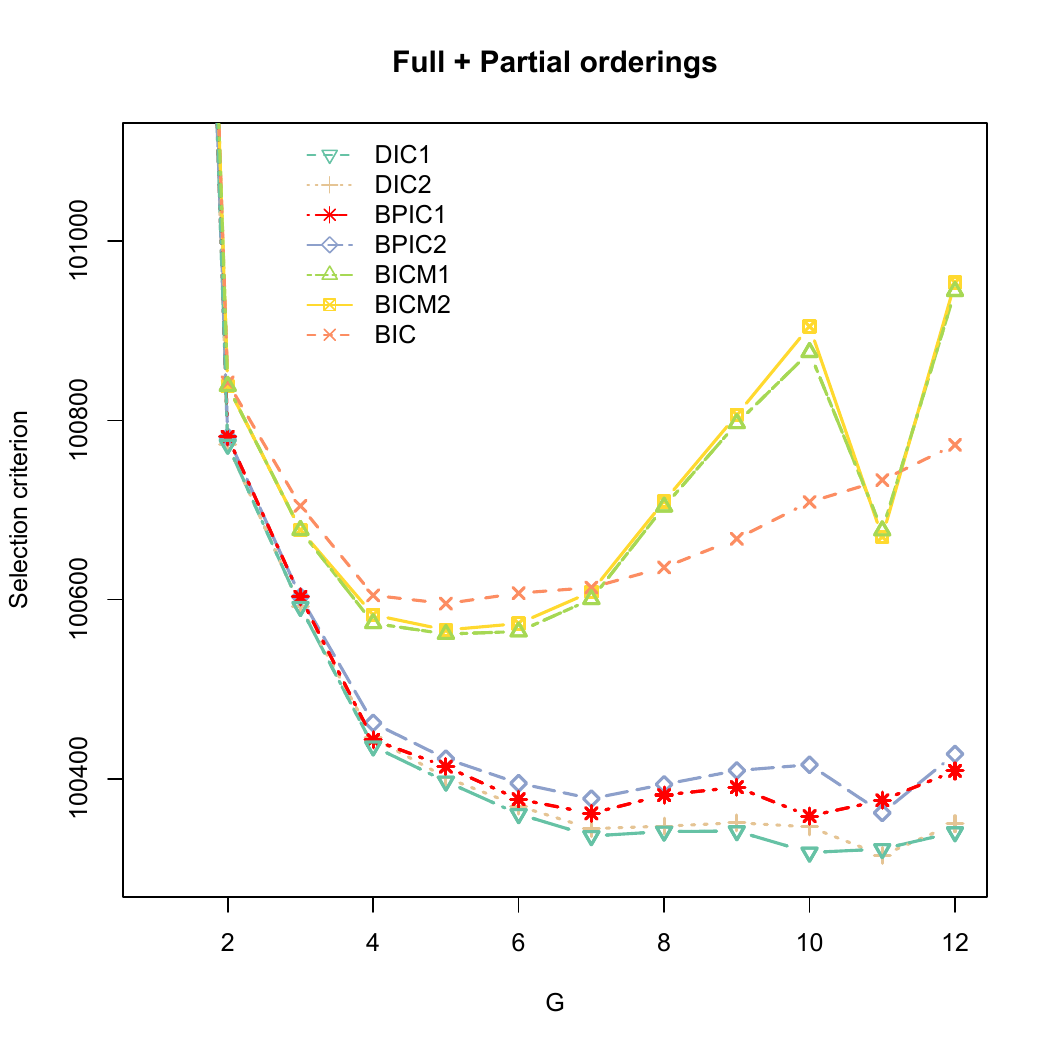}
\label{fig:APAtoptotali}}
\caption{{\em APA election data (global and separate analysis for subsets of
ballots with the same number of ranked candidates)} Model selection criteria for the Bayesian PL mixtures with a varying number $G$ of components. For each selection criterion, 
the optimal choice of the number of components corresponds 
to the minimum value.}
\label{fig:APAcriteria}
\end{figure}

\clearpage

\begin{figure}[h!]
\begin{center}
\includegraphics[bb=30 15 736 311, scale=0.4]{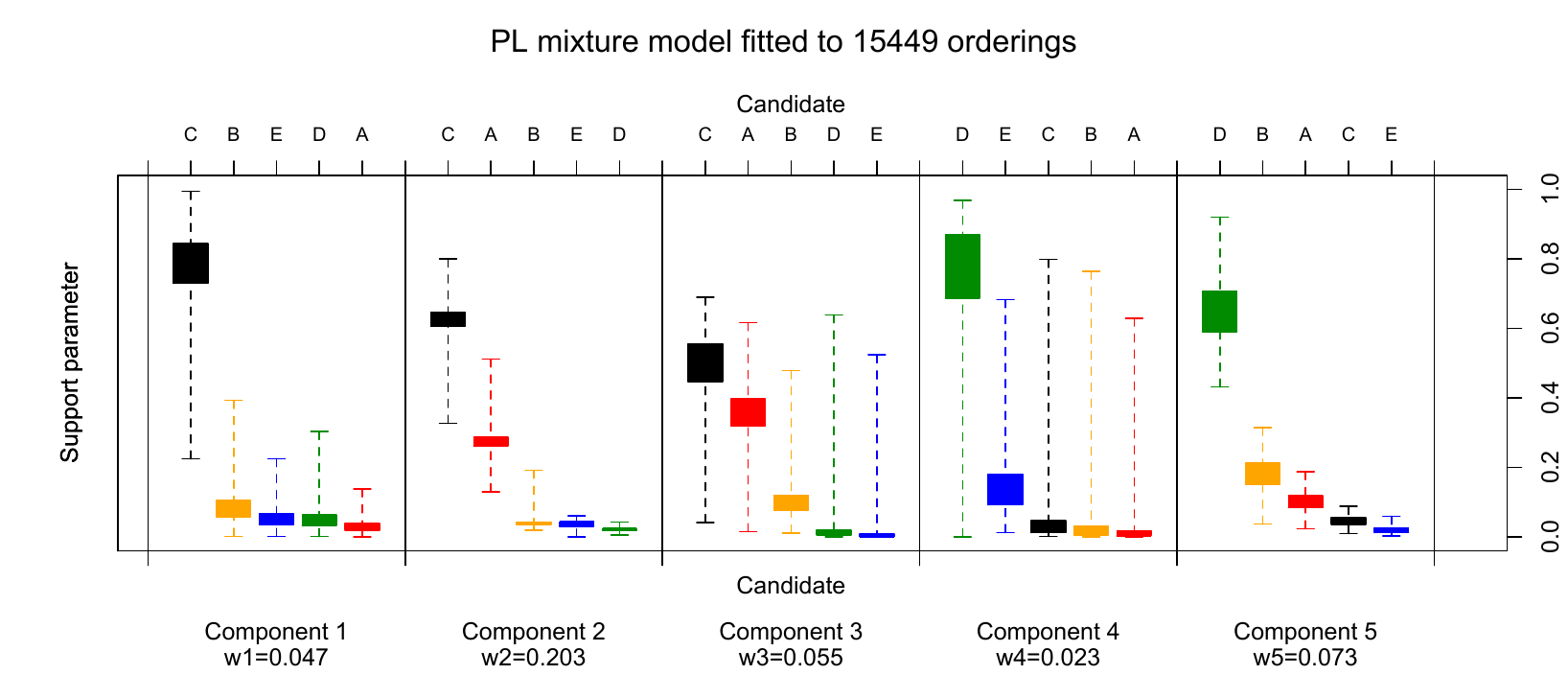}

\vspace*{0.7cm}

\includegraphics[bb=30 15 736 311, scale=0.4]{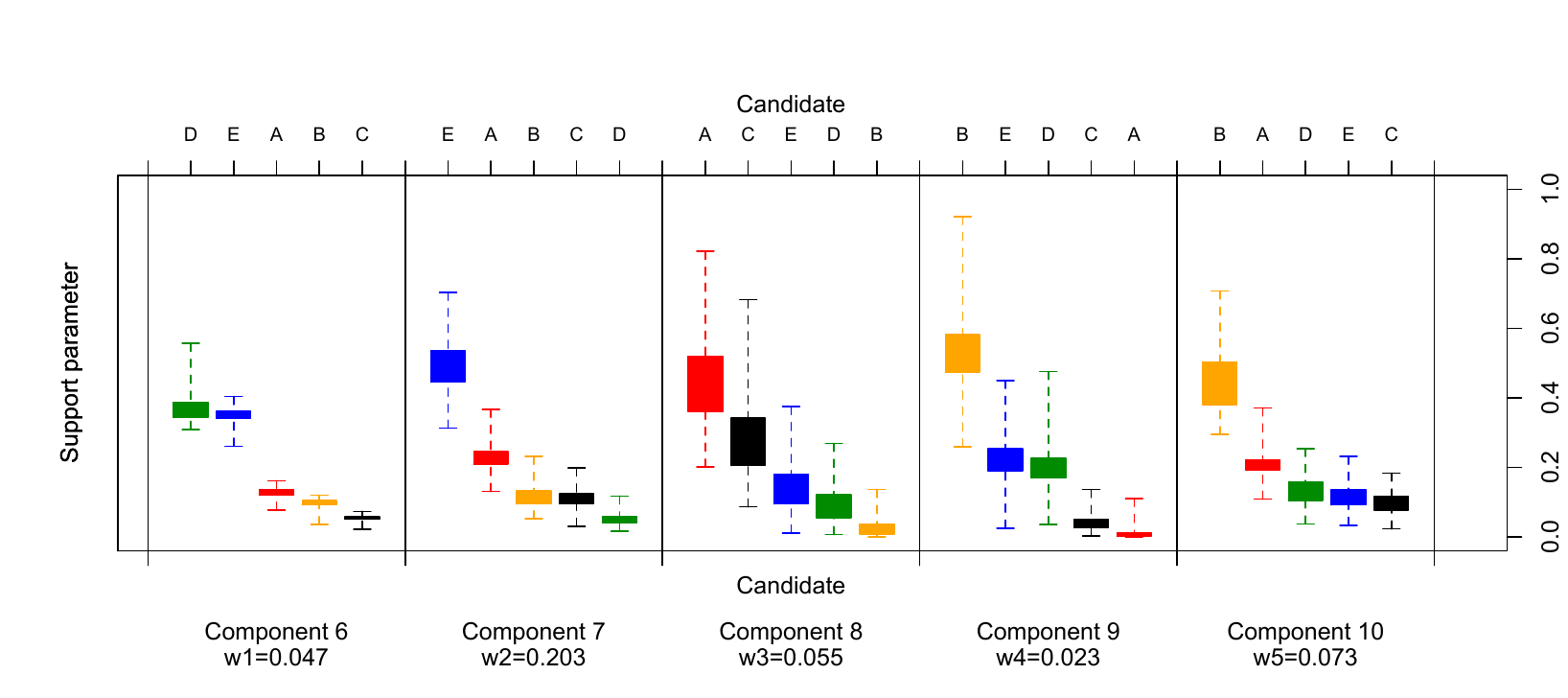}

\end{center}
\caption{{\em APA election data (global analysis with 15449 full and partial orderings)} Optimal Bayesian 10-component PL mixture model fitted to the entire data set.} 
\label{fig:global}
\end{figure}

\clearpage

\begin{figure}[h!]
\begin{center}
\includegraphics[bb=30 15 736 311, scale=0.4]{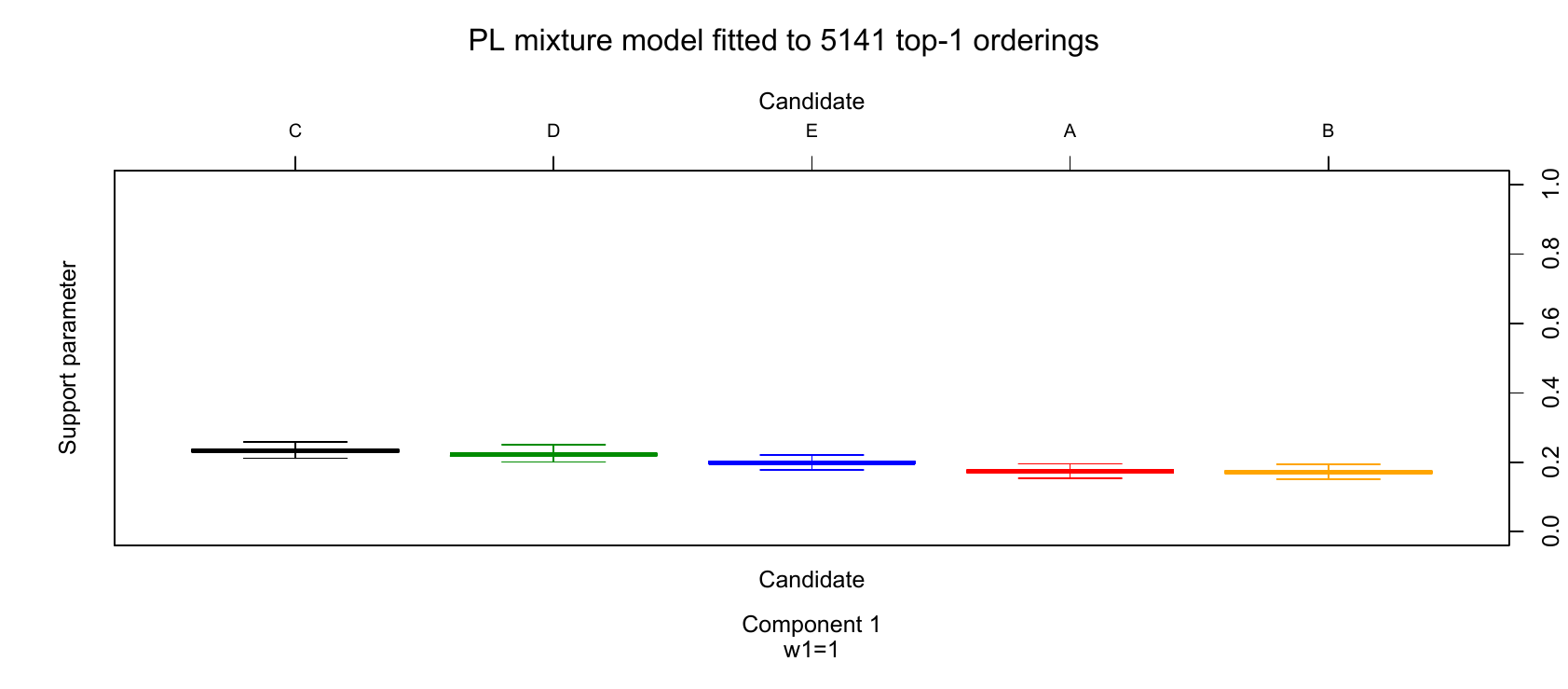}

\vspace*{0.7cm}

\includegraphics[bb=30 15 736 311, scale=0.4]{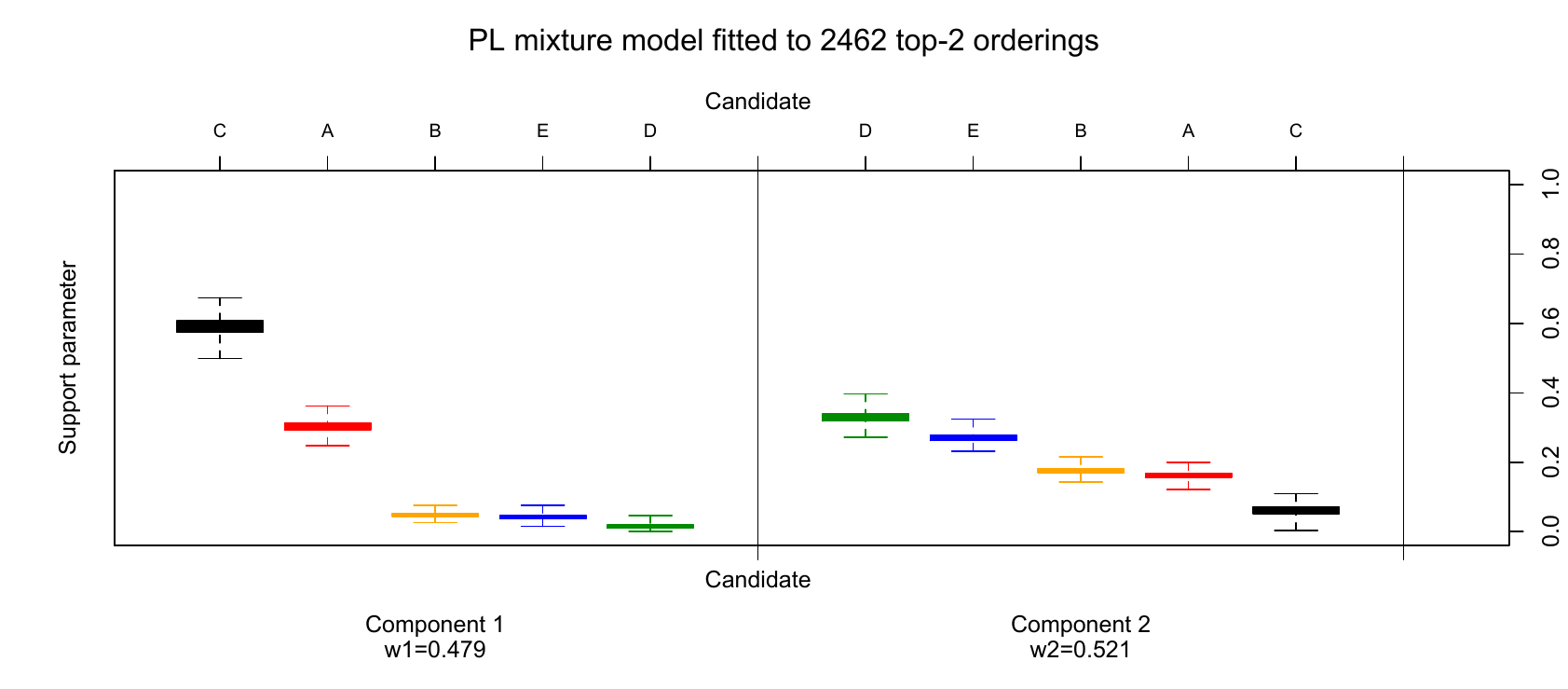}

\vspace*{0.7cm}

\includegraphics[bb=30 15 736 311, scale=0.4]{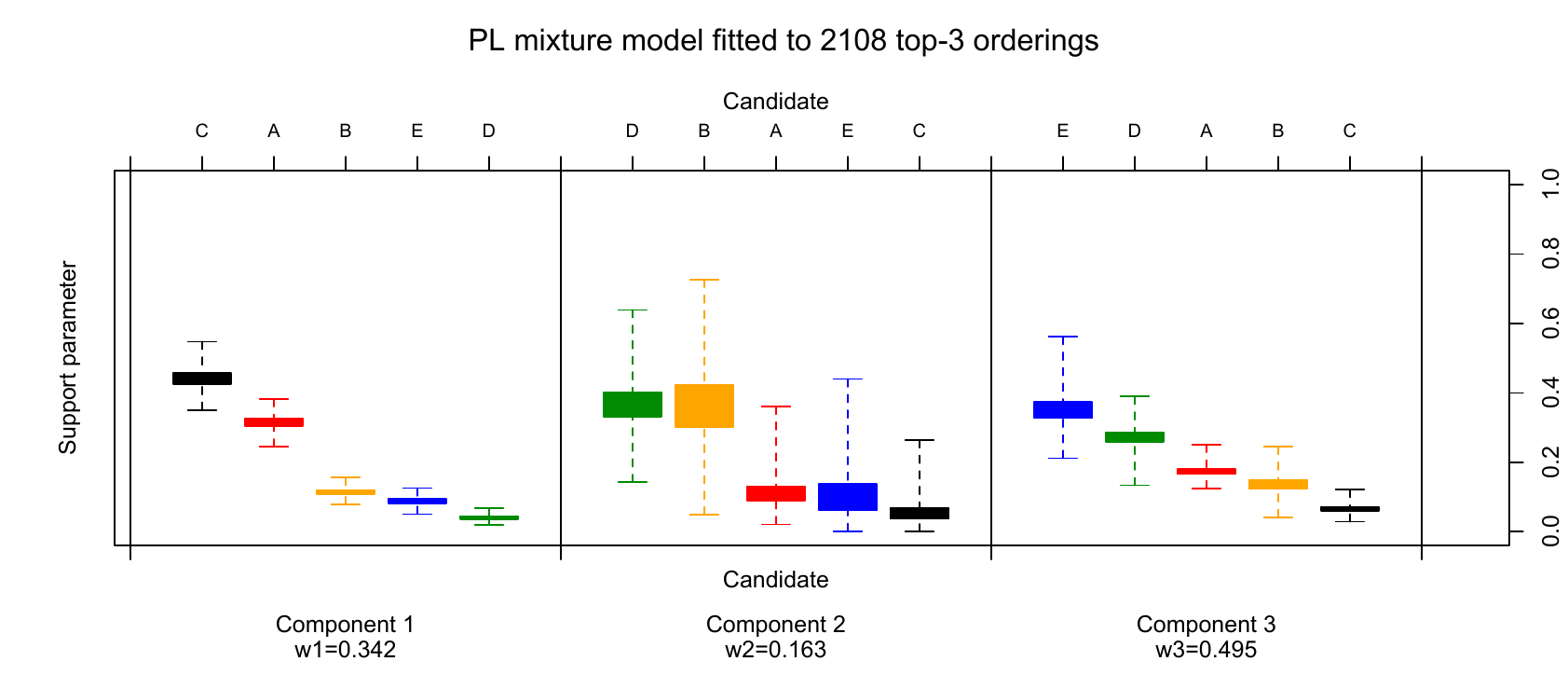}

\vspace*{0.7cm}

\includegraphics[bb=30 15 736 311, scale=0.4]{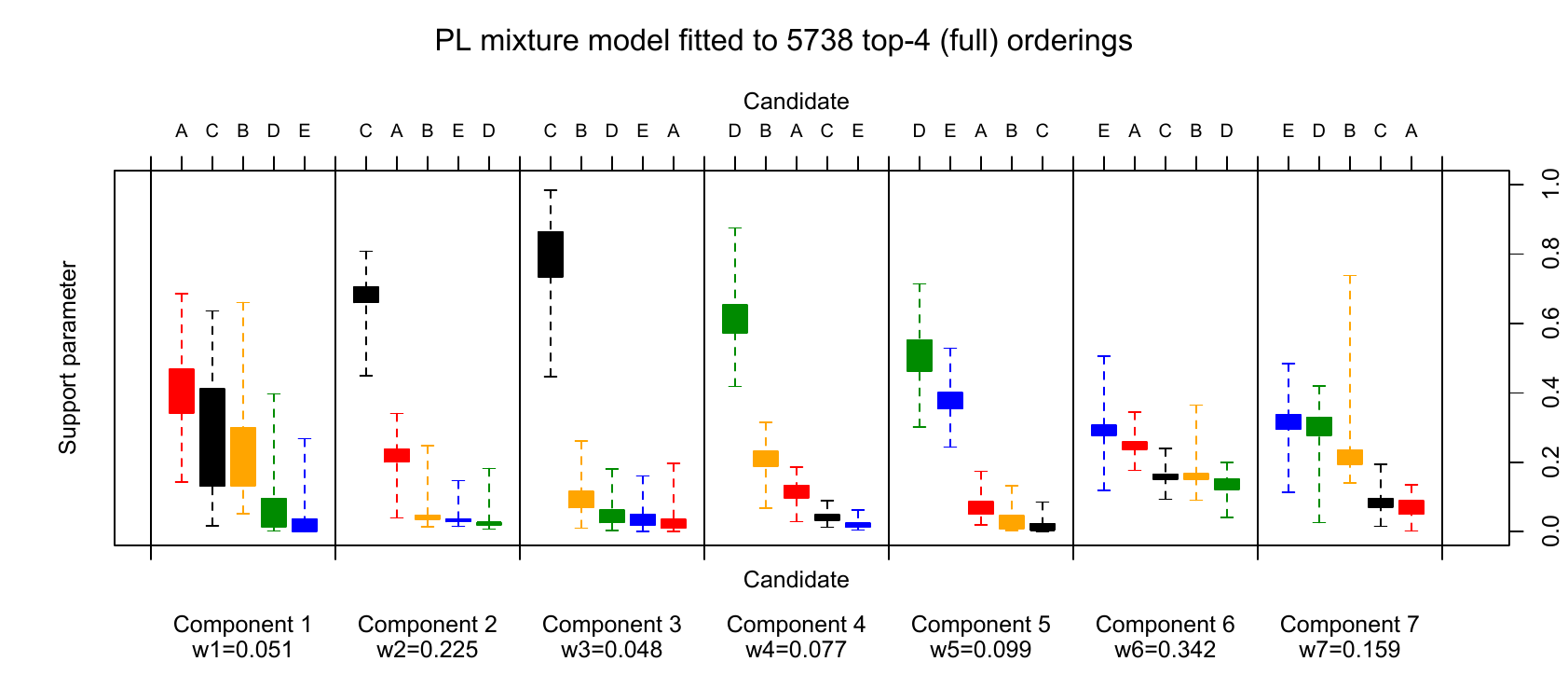}
\end{center}
\caption{{\em APA election data (separate analysis for subsets of
ballots with the same number of ranked candidates)} Optimal Bayesian PL mixtures fitted to
  subsets of partial orderings with the same number $m$ of ranked items.} 
\label{fig:separate}
\end{figure}

%
%


%


